\documentclass[twocolumn,floatfix,showpacs,pra]{revtex4}
\usepackage{amsfonts,amsmath,amssymb,amsthm}
\usepackage{graphicx,epsfig} 
 
\newtheorem{defi}{Definition}
\newtheorem{algo}{Algorithm}
\newtheorem{lemm}{Lemma}
\newtheorem{thm}{Theorem}

\begin{document}
%---------------------------------------------------------------------------
\title{The Fibonacci scheme for fault-tolerant quantum computation}

\author{Panos Aliferis$^1$ and John Preskill$^2$}
\affiliation{$^1$ IBM Watson Research Center, P.~O. Box 218, Yorktown Heights, NY 10598 \\
             $^2$ Institute for Quantum Information, California Institute of Technology, Pasadena, CA 91125
            }
%\date{\today}

\begin{abstract}
We rigorously analyze Knill's Fibonacci scheme for fault-tolerant quantum computation, which is based on the recursive preparation of Bell states protected by a concatenated error-detecting code. We prove lower bounds on the threshold fault rate of $.67\times 10^{-3}$ for adversarial local stochastic noise, and $1.25\times 10^{-3}$ for independent depolarizing noise. In contrast to other schemes with comparable proved accuracy thresholds, the Fibonacci scheme has a significantly reduced overhead cost because it uses postselection far more sparingly. 
\end{abstract}

\maketitle
%-------------------------------------------------------% 
\section{Introduction}

Scientists around the world are striving toward the construction of a scalable quantum computer. Theorists can contribute to this effort by developing and optimizing methods for protecting quantum computers against environmental decoherence and systematic hardware imperfections. Indeed, the theory of quantum fault tolerance \cite{Shor96,ref1} has established that reliable quantum computation is possible if the noise afflicting the computer has suitable properties. The most obvious requirement is that the noise must be sufficiently weak---if the noise strength is below a {\em threshold of accuracy} then quantum computing is scalable in principle. Recent insights have led to steadily improving estimates of the accuracy threshold, boosting confidence that large-scale quantum computers will be realized eventually.

Notably, Knill proposed a scheme for fault-tolerant quantum computation based on concatenated error-detecting codes and postselection, in which ancilla states are prepared off-line and an ancilla is discarded whenever an error is detected during its preparation. Numerical simulations indicate that Knill's scheme can tolerate independent stochastic noise with an error probability per gate of about $1\%$ \cite{Knill05}, and a threshold error rate of order $.1\%$ has been rigorously established \cite{Reichardt06, Aliferis07b}. Unfortunately, the overhead cost of Knill's postselection scheme is discouraging because ancillas are rarely accepted.  However, Knill also proposed a modified version of the scheme with a more moderate overhead cost, which we call the ``Fibonacci scheme.'' The Fibonacci scheme is based on the observation that a distance-2 code, aside from detecting errors, can also correct errors that occur at known positions in the code block; thus, in a concatenated code, detected errors at one coding level become {\em located} errors that can be corrected at the next level up. (We call it the Fibonacci scheme because the probability of a logical error scales with the coding level $j$ like $\varepsilon^{F(j)}$, where $\varepsilon$ is the noise strength and $F(j)$ is a Fibonacci sequence.) Knill's numerical simulations indicate that the Fibonacci scheme, too, can tolerate independent stochastic noise with an error rate of about $1\%$ \cite{Knill05}. 

In this paper, we provide a rigorous analysis of the Fibonacci scheme. Our main result is a lower bound on the accuracy threshold for the Fibonacci scheme that is of the same order as the lower bound found earlier for Knill's postselection scheme. Thus our result proves that a high accuracy threshold can be compatible with a moderate overhead cost, supporting Knill's numerical findings. 

The Fibonacci scheme has some features in common with fault-tolerant schemes used in previous proofs of the quantum threshold theorem, but there are also important differences, and correspondingly our new threshold proof has a rather different structure than these previous proofs. As in previous schemes, logical qubits processed by the computer are protected from damage using a quantum code, and encoded gates are realized by gadgets that act on the code blocks and correct the errors. But standard versions of the quantum threshold theorem are based on concatenated coding and recursive simulations. The code block of a concatenated code is constructed as a hierarchy of codes within codes---the code block at level $j$ of this hierarchy is built from logical qubits encoded at level $j{-}1$ of the hierarchy. Likewise, in a recursive simulation, fault-tolerant gadgets are constructed as a hierarchy of gadgets within gadgets---the gadgets at level $j$ are built from gadgets at level $j{-}1$. This recursive construction streamlines the analysis; if the level-1 simulation of an ideal circuit is more reliable than an unprotected ``level-0'' circuit, then because of the self-similarity of the gadgets, the level-2 simulation will be still more reliable, and so on. 

In contrast, though the Fibonacci scheme uses concatenated coding, the construction of gadgets is not strictly recursive; rather, the logical controlled-{\sc not} ({\sc cnot}) gates and encoded measurements are implemented transversally at all levels of the concatenated code. That is, a logical {\sc cnot} gate is performed by executing many fundamental {\sc cnot} gates in parallel in a single time step, and encoded measurements are performed by measuring many qubits in parallel in a single time step, followed by classical decoding of the measurement outcomes. Errors are corrected by means of quantum teleportation, using encoded Bell pairs that {\em are} prepared recursively, and the analysis of the recursive Bell pair preparation procedure is the crux of our proof. For the scheme to succeed, it is essential that the errors in an encoded Bell pair are only weakly correlated; our central task is to formalize a notion of ``weakly correlated'' that makes the proof work, and to show that the encoded Bell pairs really have this property at each level of concatenation.

A second distinctive feature of the Fibonacci scheme, as already mentioned, is that a concatenated error-{\em detecting} code is used to correct {\em located} errors. The {\em classical} decoding of a measured code block is performed recursively, starting at the lowest level; a subblock is flagged if an error is detected in the subblock, and at each level, if an error is detected in a block that has only one flagged subblock, then the flagged subblock is corrected. In principle, flagging detected errors and constructing non-recursive gadgets are independent ideas---we could use flagging in a recursive scheme or use non-recursive gadgets without flagging. However, following Knill, we have found that combining flagging and non-recursive gadget construction gives an especially high threshold estimate at a moderate overhead cost. 

This paper is structured as follows. In section \ref{sec:noise-model}, we define the correlated ``local stochastic'' noise model that we consider in most of our analysis. (Our arguments also apply to uncorrelated Pauli noise models, and yield a slightly better lower bound on the accuracy threshold in that case.) In section \ref{sec:scheme}, we formulate the Fibonacci scheme for simulating a quantum circuit, and present gadgets for error correction and logical gates based on a concatenated 4-qubit code; then in section \ref{sec:decoding} we define our procedure for decoding the measurement of a code block.  In sections \ref{sec:level-reduction} and \ref{sec:CSS-gadgets}, we derive recursion relations that characterize the error correlations in recursively prepared Bell pairs and the failure probability of gadgets for $\mathcal{G}_{\rm CSS}$ operations; this is the technical heart of our proof. In section \ref{sec:CSS-threshold}, we obtain a lower bound on the accuracy threshold for $\mathcal{G}_{\rm CSS}$ operations, and we extend it to a complete universal set of fault-tolerant quantum gates in section \ref{sec:threshold}. We discuss the overhead cost of the Fibonacci scheme in section \ref{sec:overhead}. Finally, section \ref{sec:conclusion} contains our conclusions.

%---------------------------------------------------%
\section{Noise model}
\label{sec:noise-model}

We use the term {\em location} to speak of an operation in a quantum circuit that is performed in a single time step; a location may be a single-qubit or multi-qubit gate, a qubit preparation step, a qubit measurement, or the identity operation in the case of a qubit that is idle during the time step. In a stochastic noise model, we assume that at each circuit location either the ideal operation is executed perfectly or else a fault occurs, and we assign a probability to each {\em fault path}---that is, to each possible set of faulty locations in the circuit. We speak of {\em local stochastic noise} with strength $\varepsilon$ if, for any $r$ specified locations in the circuit, the sum of the probabilities of all fault paths with faults at those $r$ locations is no larger than $\varepsilon^r$. For each fault path, the trace-preserving quantum operation applied at the faulty locations is arbitrary and can be chosen adversarially; therefore in this noise model the faults can be correlated both temporally and spatially. However, we will exclude ``leakage'' faults in which quantum information escapes from the computational space. 

In the Fibonacci scheme, errors are corrected by teleportation, and logical errors occur if the Bell measurements in the noisy circuit fail to simulate correctly the Bell measurements in the ideal circuit. We will analyze the effectiveness of the scheme by propagating errors forward to the measurements, in order to estimate the failure probability for the measurement of encoded blocks. This error propagation is feasible because qubits are always measured just a few time steps after they are initially prepared.

For each fault path, the operation $\mathcal{N}$ acting at the faulty locations can be expressed in terms of a set $\{N_k\}$ of nontrivial Kraus operators, and each Kraus operator has a Pauli expansion $N_k = \sum_l c_{k,l}\, P_l$ where $P_l$ denotes a tensor product of Pauli spin operators $\{I, \sigma_{\rm x}, \sigma_{\rm y},\sigma_{\rm z}\}$ that act at the faulty locations. Each Pauli operator can be propagated forward in time through the ideal circuit. Thus, for each fault path and for each time $\tau$, we obtain a trace-preserving completely positive map $\mathcal{N}(\tau)$ acting on the qubits at time $\tau$, where each Kraus operator in $\mathcal{N}(\tau)$ also has a Pauli expansion. For any qubit labeled $q$ we will say that the operation $\mathcal{N}$ causes a ``type-x'' error acting on $q$ at time $\tau$ if any Pauli operator in the expansion of $\mathcal{N}(\tau)$ contains either $\sigma_{\rm x}$ or $\sigma_{\rm y}$ acting on $q$; similarly we say that $\mathcal{N}$ causes a ``type-z'' error acting on $q$ at time $\tau$ if any Pauli operator in the expansion of $\mathcal{N}(\tau)$ contains either $\sigma_{\rm z}$ or $\sigma_{\rm y}$ acting on $q$. Thus, if $q$ is measured at time $\tau$, then $\mathcal{N}$ generates a measurement error with nonzero probability if the measurement is in the z basis and there is a type-x error, or if the measurement is in the x basis and there is a type-z error.

%---------------------------------------------------%
\section{Scheme}
\label{sec:scheme}

We will assume that our quantum computer is equipped to execute elementary operations chosen from the universal set
%
%\begin{widetext}
\begin{equation}
\label{eq:1}
\begin{array}{rcl}
\mathcal{G}  & = & \overbrace{\{ {\rm{\small CNOT}}, \, \mathcal{P}_{|0\rangle}, \mathcal{P}_{|+\rangle}, \mathcal{M}_{\sigma_{\rm z}}, \mathcal{M}_{\sigma_{\rm x}} \}}^{\mathcal{G}_{\rm CSS}} \vspace{.2cm} \\
             &   & \cup \, \{\mathcal{P}_{|{+}i\rangle}, \mathcal{P}_{|T\rangle} \} \; ;
\end{array}
%\vspace{.2cm}
\end{equation}
%\end{widetext}
%
\noindent here, $\mathcal{M}_{\sigma_{\rm z}}$ and $\mathcal{M}_{\sigma_{\rm x}}$ denote the measurement of the single-qubit Pauli operators $\sigma_{\rm z}$ and $\sigma_{\rm x}$ respectively and $\mathcal{P}_{|\phi\rangle}$ denotes the preparation of the single-qubit state $|\phi\rangle$, where $|+\rangle = {1\over \sqrt{2}} (|0\rangle + |1\rangle)$, $|{+}i\rangle = {1\over \sqrt{2}} (|0\rangle + i |1\rangle)$, and $|T\rangle = {1\over \sqrt{2}} (|0\rangle + e^{i\pi/4}|1\rangle)$. To prove our threshold theorem, first we will show that there is an accuracy threshold $\varepsilon^{\rm css}_0$ for operations in $\mathcal{G}_{\rm CSS}$---if $\varepsilon \leq \varepsilon^{\rm css}_0$ then {\em encoded} $\mathcal{G}_{\rm CSS}$ operations can be simulated to any desired accuracy. Then we will show that for $\varepsilon \leq \varepsilon^{\rm css}_0$, by using the reliable encoded $\mathcal{G}_{\rm CSS}$ operations and the noisy elementary operations $\mathcal{P}_{|{+}i\rangle}$ and $\mathcal{P}_{|T\rangle}$, we can distill accurately encoded copies of  $|{+}i\rangle$ and $|T\rangle$, thus completing a universal set of reliable encoded operations. 

%---------------------------------------------------%

In the Fibonacci scheme, operations in $\mathcal{G}$ will be protected using the code $(C_4)^{\triangleright \, j}$, the 4-qubit code $C_4$ concatenated $j$ times. (Knill actually proposed to concatenate  $C_4$ with a 6-qubit code $C_6$ in order to reduce the overhead \cite{Knill05}, but we will analyze a simpler scheme that uses $C_4$ only.) The code $C_4$ has distance 2 and its check operators are $\sigma_{\rm z}^{\otimes 4}$ and $\sigma_{\rm x}^{\otimes 4}$. Though the code space is four dimensional, we will use only one of the two encoded qubits to protect the quantum information processed by the computer. The encoded Pauli operators acting on this ``logical'' qubit are
\begin{equation}
\label{eq:2}
\sigma_{\rm z}^{L} = \sigma_{\rm z}\otimes I\otimes \sigma_{\rm z} \otimes I \;, \; \;  \sigma_{\rm x}^{L} = \sigma_{\rm x}\otimes \sigma_{\rm x} \otimes I \otimes I \; ,
\end{equation}
\noindent while the encoded Pauli operators acting on the irrelevant ``gauge'' qubit  are
\begin{equation}
\label{eq:3}
\sigma_{\rm z}^{G} = I\otimes I\otimes \sigma_{\rm z}\otimes \sigma_{\rm z} \;, \; \;  \sigma_{\rm x}^{G} = \sigma_{\rm x}\otimes I\otimes  \sigma_{\rm x} \otimes I \; .
\end{equation}
\noindent We do not care about the state of the gauge qubit, and therefore a $\sigma_{\rm z}$ error acting on the third qubit in the code block is equivalent to a $\sigma_{\rm z}$ error acting on the fourth qubit---the errors differ by the gauge-qubit operator $\sigma_{\rm z}^{G}$, which commutes with the logical Pauli operators. Likewise, $\sigma_{\rm z}$   errors on the first and second qubits are equivalent, as are $\sigma_{\rm x}$ errors on the first and third qubits or $\sigma_{\rm x}$ errors on the second and fourth qubits. 

\begin{figure}[t]
\begin{center}
\vspace{-.2cm} \hspace{-.1cm}
\includegraphics[width=8.3cm,keepaspectratio]{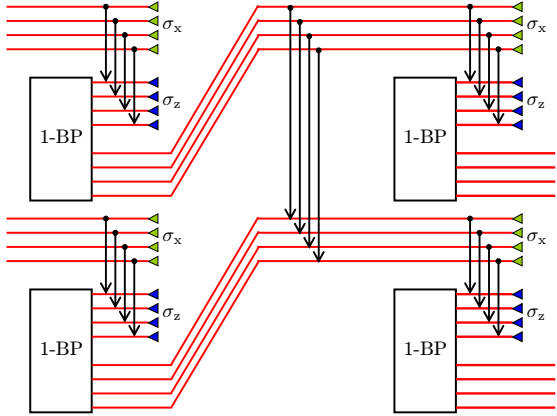}
\vspace{-1.3cm} 
\end{center}
\caption{\label{fig:1} The gadget for a {\sc cnot} gate protected by $C_4$. A transversal {\sc cnot} gate, which implements the logical {\sc cnot} gate acting on two blocks, is preceded and followed by logical teleportations which extract the code's error syndrome. Arrows denote {\sc cnot} gates, with the arrow tip pointing to the target qubit. Green and blue triangles pointing to the left denote $\mathcal{M}_{\sigma_{\rm x}}$ and $\mathcal{M}_{\sigma_{\rm z}}$ respectively, and Bell pairs encoded in $C_4$ (1-BPs) are prepared as in Fig.~\ref{fig:2}.
         }
\end{figure}

We will derive our lower bound on $\varepsilon^{\rm css}_0$ by showing that, for  $\varepsilon < \varepsilon^{\rm css}_0$, ``gadgets'' realizing encoded $\mathcal{G}_{\rm CSS}$ operations protected by $(C_4)^{\triangleright \, j}$ become highly reliable as $j$ increases. How are these gadgets constructed? First, consider the $\mathcal{G}_{\rm CSS}$ gadgets protected by $C_4$ (the case $j=1$). The {\sc cnot} gadget, shown in Fig.~\ref{fig:1}, consists of a transversal logical {\sc cnot} gate, together with error recovery steps on each block that precede and follow the transversal gate. The other $\mathcal{G}_{\rm CSS}$ gadgets are smaller and more reliable than the {\sc cnot} gadget, so to obtain a lower bound on the $\mathcal{G}_{\rm CSS}$ threshold it suffices to study the {\sc cnot} gadget. 

\begin{figure}[t]
\begin{center}
\vspace{-.5cm} \hspace{-.1cm}
\includegraphics[width=8.3cm,keepaspectratio]{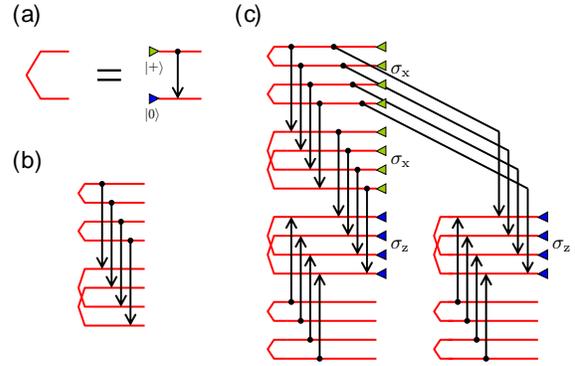}
\vspace{-2cm}
\end{center}
\caption{\label{fig:2} (a) Preparation of an unencoded Bell pair (0-BP) where green and blue triangles pointing to the right denote $\mathcal{P}_{|+\rangle}$ and $\mathcal{P}_{|0\rangle}$ respectively. (b) A Bell pair encoded in $C_4$ (1-BP) is constructed by preparing one block in the logical $|+\rangle$ state, a second block in the logical $|0\rangle$ state, and applying transversal {\sc cnot} gates. (c) The two blocks of a 1-BP are teleported, using two other 1-BPs, in order to limit correlations between errors in different output blocks. Depending on the measurement outcomes, logical Pauli operators (not shown) acting on the two output blocks may be needed to complete the teleportations. 
         }
\end{figure}

Inside all gadgets, error recovery is achieved by logical teleportation using Bell pairs encoded in $C_4$. The encoded Bell pairs (1-BPs) are prepared by using unencoded Bell pairs (0-BPs) and {\sc cnot} gates as in Fig.~\ref{fig:2}. The preparation circuit must be designed carefully to control correlated errors.  We do not need to worry about correlated errors in the 0-BPs---any two-qubit error acting on  the Bell state $|\Phi_0\rangle = {1\over \sqrt{2}}(|0\rangle \otimes |0\rangle + |1\rangle \otimes |1\rangle)$ is equivalent to a one-qubit error, because $|\Phi_0\rangle$ is invariant under $\sigma_{\rm x}\otimes \sigma_{\rm x}$ and $\sigma_{\rm z}\otimes \sigma_{\rm z}$. On the other hand, error correlations between the two blocks of the  1-BP might cause trouble, and can arise from a single fault in the transversal {\sc cnot} gate shown in Fig.~\ref{fig:2}(b). A simple way to suppress correlated errors is to follow the preparation of the 1-BP with the teleportation of both blocks, using two other 1-BPs as shown in Fig.~\ref{fig:2}(c). Using this circuit, an error occurs in an output block only if there is a fault in the teleportation of the block; therefore, errors in different blocks must arise from distinct faults.

One way to construct the gadget for the {\sc cnot} gate encoded in $(C_4)^{\triangleright \, 2}$ would be to replace every elementary {\sc cnot} gate in Fig.~\ref{fig:1} by the {\sc cnot} gadget encoded in $C_4$. And by repeating this replacement recursively, we could obtain the ``level-$j$'' gadget for the {\sc cnot} gate encoded in $(C_4)^{\triangleright \, j}$ for any $j>1$ . However, in the Fibonacci scheme we do {\em not} follow this recursive procedure to construct the higher-level gadgets. Instead, for each $j$ the logical {\sc cnot} gate inside the level-$j$ {\sc cnot} gadget is performed transversally in a single time step just as in Fig.~\ref{fig:1}, and the measurements of encoded blocks are also transversal. We construct the higher-level gadgets using this non-recursive method because we have found that it yields a significantly better lower bound on the accuracy threshold than a strictly recursive procedure. On the other hand, the level-$j$ Bell pair ($j$-BP) inside the level-$j$ {\sc cnot} gadget {\em is} constructed recursively. That is, just as the 1-BP preparation circuit in Fig.~\ref{fig:2}(c), constructed from transversal level-1 {\sc cnot} gates and measurements, operates on twelve input 0-BPs, so the $j$-BP preparation circuit, constructed from level-$j$ transversal {\sc cnot} gates and measurements, acts on twelve input $(j{-}1)$-BPs.

But this circuit for preparing a $j$-BP still has a flaw that must be addressed---for $j$ large, each output qubit of the $j$-BP goes unmeasured for many time steps, gradually accumulating errors. Let us say that the final step in the $j$-BP preparation circuit, the level-$j$ transversal {\sc cnot} gate acting on the output blocks, occurs in time step $j$. This transversal {\sc cnot} follows $(j{-}1)$-BP preparations that are completed in time step $j{-}1$. Thus, there are level-$(j{-}1)$ transversal {\sc cnot} gates in time step $j{-}1$, which are preceded by $(j{-}2)$-BP preparations that are completed in time step $j{-}2$, and so on. We see that each of the output qubits of the $j$-BP is acted on by {\sc cnot} gates in consecutive time steps 1 through $j$. But during these $j$ steps nothing is being done to correct the errors in these qubits.

\begin{figure}[t]
\begin{center}
\vspace{-.8cm} %\hspace{-.4cm}
\includegraphics[width=8.2cm,keepaspectratio]{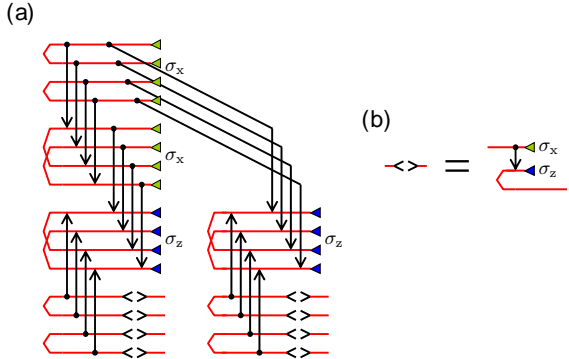}
\vspace{-1.7cm}
\end{center}
\caption{\label{fig:3} (a) The complete preparation circuit for a $j$-BP where $j>1$. It is identical to the circuit shown in Fig.~\ref{fig:2}(c), except that a final teleportation of each level-$(j{-}1)$ subblock has been added.  (b) The teleportation circuit.  
         }
\end{figure}

We can correct this flaw by adding one more ingredient to the scheme; for $j>1$ we replace the circuit in Fig.~\ref{fig:2}(c), by the circuit in Fig.~\ref{fig:3}. In this modified circuit, after a $j$-BP is prepared, we teleport every output $(C_4)^{\triangleright \, (j{-}1)}$ subblock by using additional $(j{-}1)$-BPs. In the modified $j$-BP preparation circuit, then, the last step is a $(j{-}1)$-BP preparation acting on the subblocks. In turn, the last step in the $(j{-}1)$-BP preparation is a $(j{-}2)$-BP preparation, the last step in the $(j{-}2)$-BP preparation is a $(j{-}3)$-BP preparation, and so on. Thus, the output qubits from the circuit actually emerge directly from 1-BPs, and there is no accumulation of error to worry about.

To determine the effectiveness of this scheme, where errors are corrected by logical teleportation, we need to estimate the probability of failure when a $(C_4)^{\triangleright \, j}$ code block is measured. Measurements are performed transversally---to measure the encoded $\sigma_{\rm x}$, we measure all qubits in the block simultaneously in the x basis, and to measure the encoded $\sigma_{\rm z}$, we measure all qubits in the block simultaneously in the z basis. Then these measurement outcomes are decoded using a classical recursive algorithm that is assumed to be noiseless. First, the $C_4$ subblocks are decoded, then the $(C_4)^{\triangleright \, 2}$ subblocks, {\em etc}., up to the final $(C_4)^{\triangleright \, j}$ block. At each level $i \ge1 $, the decoding of a $(C_4)^{\triangleright \, (i{+}1)}$ block makes use of the error syndrome computed in the previous step during the decoding of its four $(C_4)^{\triangleright \, i}$ subblocks. Specifically, whenever we detect an error in a $(C_4)^{\triangleright \, i}$ subblock for $1\le i<j$, we {\em flag} this subblock to indicate that the decoding of the subblock has an enhanced probability of failure. Then, if a $(C_4)^{\triangleright \, (i+1)}$ block has a nontrivial error syndrome and a single flagged subblock, we decode the $(C_4)^{\triangleright \, (i+1)}$ block under the assumption that the flagged subblock was decoded incorrectly. 

At the top level, an error might be detected in the decoding of the final $(C_4)^{\triangleright \, j}$ block that cannot be corrected (either because no $(C_4)^{\triangleright \, (j{-}1)}$ subblock is flagged, or because more than one subblock is flagged). If such an uncorrectable error is found at the top level in the recursive preparation of an encoded Bell pair, then the preparation is aborted. On the other hand, if such an uncorrectable error is detected at a measurement's top level in the teleportation step inside a gadget, then the code has failed to protect the measurement effectively. We will prove that, for sufficiently weak noise, the probability of detecting an error in the final decoding step of a $(C_4)^{\triangleright \, j}$ block decreases rapidly with increasing $j$. Thus the overhead cost due to aborting and restarting the preparation of encoded Bell pairs is moderate; the preparation of a $j$-BPs is rarely aborted once $j$ is large. Furthermore, the probability of a logical error inside a level-$j$ gadget becomes arbitrarily small for $j$ sufficiently large.

%---------------------------------------------------%
\section{Decoding $C_4$}
\label{sec:decoding}

Now we will specify in more detail the (noiseless) classical algorithm for decoding measurements of code blocks, and we will discuss the effectiveness of this algorithm. Under appropriate assumptions about the noise in the measurements, we would like to estimate how the probability of error and the probability of flagging propagates from level to level as the block is recursively decoded. 

At each level of the recursive decoding, we are to interpret a 4-bit string corresponding to the measured eigenvalues of the encoded $\sigma_{\rm x}$ or the encoded $\sigma_{\rm z}$ in the four $(C_4)^{\triangleright \, i}$ subblocks of a $(C_4)^{\triangleright \, (i+1)}$ block, where some of the bits might be flagged. The decoder knows which bits are flagged, but of course it does not know which bits have errors. We may consider a joint distribution that governs the probability of error and the probability of flagging for the four bits. For each measurement in the Fibonacci simulation, and for each level of the concatenated code, this distribution is completely determined in principle by the decoding procedure at lower levels and by the noise model. That is, if we know the probability of each fault path and the noise operation associated with each fault path, we can propagate the errors forward (as described in section \ref{sec:noise-model}) to determine an error model for the measured block. Then using this error model and the decoding procedure, we can assign a probability (or at least an upper bound on the probability) for each nontrivial pattern of flagging and errors at each decoding level.

To prove our threshold theorem, we will need to show that the decoding errors at each concatenation level are only weakly correlated. A suitable notion of weakly correlated, which we call ``quasi-independence,'' is defined as follows:

\begin{defi}[Quasi-independent errors] \label{def:1} Let $P(u,v)$ be a probability distribution for a set $\mathcal{I}$ of $n$ bits, where the $n$-bit binary vector $u$ specifies the positions of bits with errors, and the $n$-bit binary vector $v$ specifies the positions of flagged bits. Let $\mathcal{J}_r$ and $\mathcal{K}_s$ denote disjoint subsets of $\mathcal{I}$ containing $r$ bits and $s$ bits respectively, and let $\mathcal{L}_t$ denote a subset of $\mathcal{K}_s$ containing $t$ bits. We say that the distribution $P(u,v)$ is {\em quasi-independent} with strength $\{ f, \delta_{f},\delta_{\neg f}\}$ if, for any choice of $\mathcal{J}_r$, $\mathcal{K}_s$, and $\mathcal{L}_t$, the probability that all bits in $\mathcal{J}_r$ are unflagged with errors, all bits in $\mathcal{K}_s$ are flagged, and all bits in $\mathcal{L}_t$ are flagged with errors is no larger than $(\delta_{\neg f})^r  (\delta_{f})^t f^{s-t}$. We say that $\delta_{\neg f}$ is the error strength for unflagged bits,  $\delta_{f}$ is the error strength for flagged bits, and $f$ is the flagging strength.
\end{defi} 
\noindent When applied to the decoding of $\sigma_{\rm z}^L$, the errors of interest are $\sigma_{\rm x}$ errors; hence quasi-independence constrains how $\sigma_{\rm x}$ errors are correlated with one another. Likewise, when applied to the decoding of $\sigma_{\rm x}^L$, quasi-independence constrains how the $\sigma_{\rm z}$ errors are correlated with one another. But we will not need to know much about how $\sigma_{\rm x}$ errors are correlated with $\sigma_{\rm z}$ errors to analyze the failure probability for measurements of encoded blocks.

Now we would like to show that when we recursively decode a $(C_4)^{\triangleright \, j}$ block, if the errors are quasi-independent at level $i$ they will also be quasi-independent at level $i{+}1$; furthermore, we wish to estimate the error strengths at level $i{+}1$ given an estimate of the error strengths at level $i$. First, let us formulate precisely our algorithm for flagging a $C_4$ block: 

%\begin{table*}
%\begin{center}
%\begin{tabular}{l}
%\hline \hline 
\begin{algo}
%{\sc Algorithm for flagging a $C_4$ block} \\ \hline 
%\begin{minipage}{16cm} \begin{flushleft}
$(a)$ If no bit in the block is flagged, then we flag the block whenever the syndrome is non-trivial, 
%\end{flushleft} \end{minipage} \vspace{.0cm} \\
%\begin{minipage}{16cm} \begin{flushleft}
$(b)$ If exactly one bit in the block is flagged, then we {\em never} flag the block, 
%\end{flushleft} \end{minipage} \vspace{.1cm} \\
%\begin{minipage}{16cm} \begin{flushleft}
$(c)$ If two or more bits in the block are flagged, then we flag the block if errors on a pair of flagged bits can cause a logical error. 
%$(c)$ If two or more bits in the block are flagged, then we flag the block if either (1) the syndrome is nontrivial or (2) the syndrome is trivial and errors on a pair of flagged bits can cause a logical error. 
%\end{flushleft} \end{minipage} \vspace{.1cm} \\
%\hline \hline
%\end{tabular}
%\end{center}
%\end{table*}
\end{algo}

\noindent In case $(a)$, we flag the block because we have detected an error that we do not know how to correct. In case $(b)$, rather than flagging the block, we perform error correction (in the event of a nontrivial syndrome) under the assumption that the flagged bit has an error.  In case $(c)$, we do not flag the block if errors at the flagged bits would affect only the gauge qubit, not the logical qubit. 

Given this flagging algorithm and assuming quasi-independent errors on the bits, we can see that the $C_4$ blocks also have quasi-independent errors. We may define a function, which we call {\tt DECODE}, that maps upper bounds $\{ f^{({\rm in})}, \delta^{({\rm in})}_{f}, \delta^{({\rm in})}_{\neg f}  \}$  on the error strengths for the bits to upper bounds $\{ f^{({\rm out})} , \delta^{({\rm out})}_{f} , \delta^{({\rm out})}_{\neg f} \}$ on the error strengths for the blocks. This function is specified in Table \ref{func:decode}. In the rest of this section, we explain how the function {\tt DECODE} is derived.

%----------------------------------------------------%
\begin{table*}
\begin{center}
\begin{tabular}{l}
\hline \hline
{\bf Function} {\tt DECODE}$\;:\{ f^{({\rm in})}, \delta^{({\rm in})}_{f}, \delta^{({\rm in})}_{\neg f}  \} \mapsto  \{ f^{({\rm out})} , \delta^{({\rm out})}_{f} , \delta^{({\rm out})}_{\neg f} \}$ \\ \hline

\begin{minipage}{16cm} \begin{flushleft} 

\vspace{.1cm}
If the bits in a set of $C_4$ blocks have quasi-independent errors, then the blocks also have quasi-independent errors. The function {\tt DECODE} maps upper bounds on the error strengths $\{ f^{({\rm in})}, \delta^{({\rm in})}_{f}, \delta^{({\rm in})}_{\neg f}  \}$ for the bits to upper bounds on the error strengths $\{ f^{({\rm out})} , \delta^{({\rm out})}_{f} , \delta^{({\rm out})}_{\neg f} \}$ for the blocks.

\end{flushleft} \end{minipage} \\ \hline

\begin{minipage}{16cm} \begin{flushleft}

\vspace{.1cm}
With our algorithm for flagging a $C_4$ block, the probability that a block is {\em flagged} is  no larger than
\begin{equation}
\label{eq:func:1}
f^{({\rm out})} = 4 \delta^{({\rm in})}_{\neg f} + 4 \left( f^{(\rm in)} \right)^2 \; , 
\end{equation}

\noindent the joint probability that a block is flagged {\em and} there is a decoding error is no larger than 
\begin{equation}
\label{eq:func:2}
\delta^{({\rm out})}_{f} = 2 \delta^{({\rm in})}_{\neg f} + 4 f^{({\rm in})} \delta^{({\rm in})}_f + 4 \left( f^{({\rm in})} \right)^2 \left( 2\delta_{\neg f}^{({\rm in})} + 3 \delta_f^{({\rm in})} \right) \; , 
\end{equation}
and the joint probability that the block is {\em not} flagged {\em and} there is a decoding error is no larger than
\begin{equation}
\label{eq:func:3}
\delta^{({\rm out})}_{\neg f} = 4 \left( \delta^{({\rm in})}_{\neg f} \right)^2 + 8 f^{({\rm in})} \delta^{({\rm in})}_{\neg f}  \; . \vspace{.3cm}  
\end{equation} 

\end{flushleft} \end{minipage} \\ \hline \hline
\end{tabular}
\end{center}
\vspace{-.4cm}
\caption{\label{func:decode} Upper bounds on error probabilities for the decoding of a $C_4$ block. 
        }
\end{table*}

Consider first the upper bound in Eq.~(\ref{eq:func:1}) on the probability of flagging a $C_4$ block. According to our algorithm, a block is flagged in the two cases $(a)$ and $(c)$. In case ($a$), an error on at least one bit in the block is necessary for the syndrome to be non-trivial. Since all bits are unflagged, the probability of flagging the block is at most $4 \delta^{({\rm in})}_{\neg f}$. In case ($c$), only 4 out of the ${4 \choose 2} = 6$ possible ways to choose two flagged bits in the block can result in a logical error. For example, if $\sigma_{\rm z}^L$ is measured, then we do not flag the block if only the first and third bits are flagged, or if only the second and fourth bits are flagged, because the weight-2 errors $\sigma_{\rm x} \otimes I \otimes \sigma_{\rm x} \otimes I$ and $I\otimes \sigma_{\rm x} \otimes I \otimes \sigma_{\rm x} $ affect only the gauge qubit, not the logical qubit. Thus, the probability of flagging the block is at most $4 ( f^{(\rm in)} )^2$. 

Consider next the upper bound in Eq.~(\ref{eq:func:2}) on the joint probability that the $C_4$ block is flagged and that decoding results in a logical error. Again we distinguish the two cases $(a)$ and $(c)$ of our flagging algorithm. In case ($a$), there is no flagged bit to help us interpret a non-trivial syndrome; by convention, then, we respond to a nontrivial syndrome by flipping the first bit in the block. Suppose that $\sigma_{\rm z}^L$ is measured. Then a logical error can arise from a $\sigma_{\rm x}$ error on the second or the fourth bit but not from a $\sigma_{\rm x}$ error on the third bit, because $\sigma_{\rm x} \otimes I \otimes \sigma_{\rm x} \otimes I = \sigma_{\rm x}^G$ commutes with $\sigma_{\rm z}^L$. A similar observation applies to a measurement of $\sigma_{\rm x}^L$. Thus, the probability of a logical error in case ($a$) is at most $2\delta_{\neg f}^{({\rm in})}$. 

We divide case ($c$) into two subcases ($c'$) and ($c''$) depending on whether there is an erroneous unflagged bit in the block or not. The probability of a logical error in case $(c')$ is at most $4 ( f^{({\rm in})} )^2 \cdot 2\delta_{\neg f}^{({\rm in})}$---at least two bits in the block must be flagged, which can be chosen in 4 ways, and with a flagged pair of bits fixed, there are at most two unflagged bits where an error may occur. 

Alternatively, in case ($c''$) there are no errors on unflagged bits. If exactly two bits are flagged, then an error on one of the two flagged bits causes a logical error, while an error on the other flagged bit does not. (In particular, if the syndrome is nontrivial, then we recover by flipping one of the two flagged bits, which is chosen by convention; a logical error occurs only if the {\em other} flagged bit has an error.) Since there are four ways to choose a pair of flagged bits, the probability of a logical error is at most $4 f^{({\rm in})} \delta_{f}^{({\rm in})}$. Otherwise, at least three bits are flagged, and for a logical error to occur, at least one of the flagged bits has an error. There are four ways to choose a set of three bits, and for each choice of three bits, there are three ways to choose which bit has the error. Therefore, the probability of a logical error is at most $4(f^{({\rm in})} )^2 \cdot 3\delta_f^{({\rm in})}$. 

Finally, consider the upper bound in Eq.~(\ref{eq:func:3}) on the joint probability that the $C_4$ block is {\em not} flagged and that decoding results in a logical error. If no bit in the block is flagged, then the syndrome must be trivial; otherwise the block would have been flagged. Therefore an even number of bits have errors, and since a weight-4 error acts trivially on the block, we may assume there are two bits with errors. Thus, the probability of a logical error is at most $4 ( \delta_{\neg f}^{({\rm in})} )^2$, as there are four pairs of bits where errors can cause a logical error. 

Alternatively, at least one bit in the block is flagged. First, suppose there is exactly one flagged bit. Then if the flagged bit has an error and there are no other errors in the block, the flagged error will be corrected successfully. We conclude therefore that, if there is a logical error, there must be an unflagged error in the block. 

The last possibility is that there are two flagged bits in the block (if there were more flagged bits we would have flagged the block), such that errors on the pair of flagged bits do not cause a logical error. In this case, too, a logical error occurs only if there is an unflagged error in the block. For suppose there is no unflagged error. Then if the syndrome is trivial, there must be either zero of two flagged errors, and either way there is no logical error. If the syndrome is nontrivial then one flagged bit has an error and the other does not. To decode, we flip one of the flagged bits (chosen by convention) in response to the nontrivial syndrome. The bit with the error might be the flipped bit or it might be the other flagged bit; either way there is no logical error.

In fact, in all cases where there is at least one flagged bit and there is a logical error, there must be a pair of bits such that one bit in the pair is flagged, the other bit has an unflagged error, and errors on the pair of bits cause a logical error. Therefore, the probability of a logical error is at most $8 f^{({\rm in})} \delta_{\neg f}^{({\rm in})}$---there are four ways to choose a pair of bits such that errors on that pair cause a logical error, and either of the two bits in the pair could be the flagged bit. This completes our explanation of the function {\tt DECODE}.

Now that the function {\tt DECODE} has been specified, we can explain how the Fibonacci scheme gets its name. Schematically, Eqs.~(\ref{eq:func:1}) and (\ref{eq:func:3}) say that the probability of flagging $f^{(j)}$ and the joint probability $\delta^{(j)}_{\neg f}$ of no flag and a logical error scale with the level $j$ of concatenation according to 
\begin{equation}
\label{flag-strength-scaling}
\begin{array}{c}
\delta^{(j)}_{\neg f} = O\left(f^{(j{-}1)}\cdot\delta^{(j{-}1)}_{\neg f}\right) \; , \vspace{.2cm} \\
          f^{(j{-}1)} = O\left(\delta^{(j{-}2)}_{\neg f}\right) \, ;
\end{array}
\end{equation}

\noindent combining these relations we find that
\begin{equation}
\delta^{(j)}_{\neg f}= O\left(\delta_{\neg f}^{(j{-}1)}\cdot \delta_{\neg f}^{(j{-}2)}\right)~.
\end{equation}
Therefore, if $\delta^{(j)}_{\neg f}$ scales with the fundamental noise strength $\varepsilon$ according to
\begin{equation}
\label{unflagged-error-scaling}
\delta^{(j)}_{\neg f} = O\left(\varepsilon^{F(j+1)}\right)~,
\end{equation}
we find
\begin{equation}
F(j{+}1)=F(j)+F(j{-}1)~,
\end{equation}
the recursion relation for a Fibonacci sequence. We can solve this recursion relation using the initial data $F(1)=1$, $F(2)=2$. Likewise, Eq.~(\ref{eq:func:2}) implies that the joint probability $\delta^{(j)}_{f}$ of flagging and a logical error scales as
\begin{equation}
\label{flagged-failure-scaling}
\delta^{(j)}_{f}= O\left(\delta^{(j-1)}_{\neg f}\right)=O\left(\varepsilon^{F(j)}\right)~.
\end{equation}
When $\varepsilon$ is small, then, the probability of a decoding failure at level $j$ is ``double-exponentially'' suppressed by a power $F(j)$ of $\varepsilon$ that grows with $j$ like a Fibonacci number: 
%
%\begin{widetext}
\begin{equation}
%F(3)=3,\;\; F(4)=5,\;\; F(5)=8, 
F(j)\sim \frac{\phi^{j{+}2}}{\phi+2} \, ,
%\vspace{.2cm}
\end{equation} 
%\end{widetext}
%
\noindent where $\phi=\frac{1}{2}\left(1+\sqrt{5}\right)\approx 1.618$ is the golden ratio.

%---------------------------------------------------%
\section{Level reduction for Bell pair preparation}
\label{sec:level-reduction}

Fault-tolerant quantum computation based on a recursive simulation can be analyzed using a procedure called ``level reduction'' \cite{Aliferis05b,Aliferis07b,Aliferis07}. In this procedure, one identifies an effective noise model that acts on quantum information encoded at level $j$ of a concatenated quantum code, and the effective noise strength $\varepsilon^{(j)}$ at level $j$ is expressed in terms of the effective noise strength $\varepsilon^{(j{-}1)}$ at level $j{-}1$. The threshold theorem is  proved by showing that $\varepsilon^{(j)}$ falls steeply as $j$ increases, provided that the fundamental ``level-0'' noise strength $\varepsilon$ is small enough.

Level reduction as practiced in \cite{Aliferis05b,Aliferis07b,Aliferis07} does not apply directly to the Fibonacci scheme, which is not strictly recursive, but we can use a modified version of the level reduction concept. The central task of our proof is to characterize the errors in recursively prepared $j$-BPs, and we will show that these errors admit a hierarchical decomposition. That is, we will identify an ``error strength'' for the errors acting on the elementary qubits in the output $(C_4)^{\triangleright \, j}$ blocks of the noisy preparation circuit of a $j$-BP, and also for the logical errors acting on the $(C_4)^{\triangleright \, i}$ subblocks for $1\le i\le j$. By an inductive argument, we will obtain an upper bound on the error strength at each level. 

The task of bounding the error strength at each level is fairly manageable because in the preparation of a $j$-BP the output level-$j$ blocks are teleported, as are the level-$(j{-}1)$ subblocks for $j\ge 2$. Because of the subblock teleportation, the noise in a $j$-BP at levels $i=0,1,\dots,j{-}2$ is the same as the noise in a $(j{-}1)$-BP at levels $i=0,1,\dots,j{-}2$ (up to a small correction due to postselection). To characterize the noise in a $j$-BP at level $j{-}1$ we need to estimate the probability of a level-$(j{-}1)$ logical error in the subblock teleportation, and to characterize the noise at level $j$ we need to estimate the probability of a level-$j$ logical error in the block teleportation. 

Such logical errors occur when the recursively decoded measurement of a level-$(j{-}1)$ subblock or a level-$j$ block in the noisy circuit disagrees with the corresponding measurement in the ideal circuit. We can estimate the probability of a measurement error for a subblock or block by repeatedly using the function {\tt DECODE} constructed in section \ref{sec:decoding}. Starting at level 0, the level-0 errors in the $(j{-}1)$-BPs and the level-0 errors due to faults in the {\sc cnot} gates are propagated forward to the qubit measurements, and the probability of a level-1 decoding error is estimated using {\tt DECODE}. Next the level-1 errors in the $(j{-}1)$-BPs are propagated forward and combined with the level-1 decoding errors; then the probability of a level-2 decoding error is estimated using {\tt DECODE}. After $j{-}1$ steps, we obtain an estimate of the probability of a logical error in the level-$(j{-}1)$ subblock teleportation. Similarly, by propagating errors at each level through the $j$-BP preparation circuit, and using {\tt DECODE} repeatedly, we can estimate the probability of a logical error in the level-$j$ block teleportation. We also need to take into account that a $j$-BP is rejected if an error is detected at the top decoding level, in either the subblock teleportation or the block teleportation. The rest of this section explains these estimates in detail.

%---------------------------------------------------%
\subsection{Noise in 0-BPs}
\label{sec:0-BPs}

\begin{figure}[t]
\begin{center}
\begin{tabular}{c}
\put(-2.5,0){\includegraphics[width=9cm,keepaspectratio]{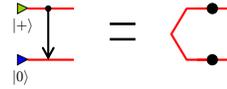}}
\vspace{-5.5cm}
\end{tabular}
\end{center}
\caption{\label{fig:4} The noisy preparation of a 0-BP is equivalent to an ideal 0-BP followed by noisy storage locations acting on the two qubits (denoted by black dots). Noise is quasi-independent and the strength for errors of type m is $\varepsilon_{\sigma_{\rm m}}(0,0)$. 
         }
\end{figure}

To begin, consider a noisy 0-BP which is prepared by the circuit in Fig.~\ref{fig:2}(a). To characterize the errors in the 0-BP, it is useful to consider a fictitious effective noisy circuit, denoted $\mathfrak{P}_{\rm 0-BP}$ and shown in Fig.~\ref{fig:4}. The effective circuit $\mathfrak{P}_{\rm 0-BP}$ consists of an ideal 0-BP preparation followed by noisy ``storage locations'' acting on the two qubits, where the faults at these noisy locations are chosen so that the output of the effective circuit matches the output of the actual noisy circuit.

Now recall that we say a fault is of ``type x'' if one of its Kraus operators does not commute with $\sigma_{\rm z}$ and of ``type z'' if one of its Kraus operators does not commute with $\sigma_{\rm x}$. The code $C_4$ is a CSS code that detects type-x errors and type-z errors separately, so to analyze the performance of the code we do not need to know how type-x errors are correlated with type-z errors. Furthermore, since the Bell state $|\Phi_0\rangle$ is left invariant by $\sigma_{\rm x}\otimes \sigma_{\rm x}$ and $\sigma_{\rm z}\otimes \sigma_{\rm z}$, any weight-two error in the effective circuit is equivalent to a weight-one error. 

Suppose that the noise in the actual circuit is local stochastic noise with strength $\varepsilon$, and consider the type-x errors. With suitable conventions, the noise in the effective circuit (or equivalently in the noisy Bell pair) is quasi-independent with strength
\begin{equation}
\label{eq:10}
\varepsilon_{\sigma_{\rm x}}(i{=}0,j{=}0) \leq \varepsilon \; ;
\end{equation} 
here, anticipating our discussion of the errors in higher-level BPs, $j=0$ indicates that we are talking about a level-0 Bell pair, and $i=0$ indicates that we are considering level-0 ({\em i.e.}, qubit) errors. Naively, one might have expected the right-hand side of Eq.~(\ref{eq:10}) to be $2\varepsilon$ rather than $\varepsilon$, because an error could arise in either a preparation step or the {\sc cnot} gate. However, a $\sigma_{\rm x}$ error acting right after the preparation $\mathcal{P}_{|+\rangle}$ of the first qubit acts trivially, while a $\sigma_{\rm x}$ error acting on the target qubit after the {\sc cnot} gate is equivalent to a $\sigma_{\rm x}$ error acting on the control qubit. A similar remark applies to $\sigma_{\rm z}$ errors, but with the target and control qubits interchanged; thus the type-z noise is also quasi-independent, with strength
\begin{equation}
\label{eq:10-typez}
\varepsilon_{\sigma_{\rm z}}(i{=}0,j{=}0) \leq \varepsilon \; .
\end{equation} 

%---------------------------------------------------%
\subsection{Noise in 1-BPs}
\label{sec:1-BPs}

\begin{figure}[t]
\begin{center}
\begin{tabular}{c}
\put(-3,0){\includegraphics[width=9cm,keepaspectratio]{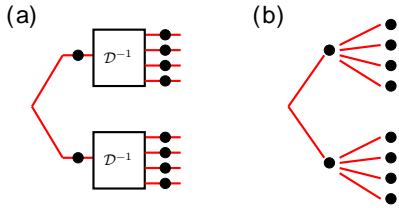}}
\vspace{-4cm}
\end{tabular}
\end{center}
\caption{\label{fig:5} (a) The effective circuit $\mathfrak{P}_{\rm 1-BP}$ that produces the same output as the 1-BP preparation circuit in Fig.~\ref{fig:2}(c); here, $\mathcal{D}^{-1}$ denotes an ideal encoder and the black dots are noisy storage locations. (b) The circuit in (a) can be represented as a directed depth-2 tree, with noisy locations at the vertices. The noise is quasi-independent for any legitimate set of vertices (a set is ``legitimate'' if it contains no descendant of a member of the set).
}
\end{figure}

Next, consider a noisy 1-BP which is prepared by the circuit in Fig.~\ref{fig:2}(c). As before, we characterize the errors in the 1-BP using a fictitious effective noisy circuit, denoted $\mathfrak{P}_{\rm 1-BP}$ and shown in Fig.~\ref{fig:5}(a), that produces the same output as the actual noisy circuit; here, $\mathcal{D}^{-1}$ denotes an ideal $C_4$ encoder, an isometry mapping the state of a single qubit to the corresponding state of the logical qubit in a $C_4$ block. There are two kinds of noisy storage locations in the effective circuit, indicated by dots---``level-1 noise locations'' that immediately precede an ideal encoder and ``level-0 noise locations'' that immediately follow an ideal encoder.

We consider separately the noise of type x and type z, and we want to formalize the idea that the noise of each type is quasi-independent at each level. We will do that by formulating and demonstrating properties of a noise model that describes the faults in the effective circuit, first for the preparation of a 1-BP, then for the preparation of a $j$-BP with $j>1$. Thus, in our inductive arguments, we will be able to justify replacing the actual noisy preparation circuit of a $j$-BP by an effective preparation circuit governed by a suitable noise model.

We will show that the noise in the effective preparation circuit is quasi-independent according to the following definition:
\begin{defi}[Quasi-independent noise] \label{def:2} Let $\mathcal{I}=\{\ell_1,\ell_2,\dots, \ell_n\}$ be a set of $n$ locations in a noisy quantum circuit. We say that the noise in $\mathcal{I}$ is {\em type-x quasi-independent with strength $\{\varepsilon_1,\varepsilon_2, \dots, \varepsilon_n\}$} if for any subset $\mathcal{J}=\{\ell_{j_1}, \ell_{j_2},\dots \ell_{j_r}\}$ of $\mathcal{I}$ the probability that there are type-{\rm x} faults at all locations in $\mathcal{J}$ is no larger than $\varepsilon_{j_1}\varepsilon_{j_2} \cdots \varepsilon_{j_r}$. Type-{\rm z} quasi-independence is defined similarly. We say that the noise in $\mathcal{I}$ is {\em fully quasi-independent} {\rm (}or, more briefly, {\em quasi-independent}{\rm )} if it is both type-{\rm x} and type-{\rm z} quasi-independent.
\end{defi}
\noindent The definition of quasi-independent noise is similar to the definition of local stochastic noise, but with three differences: it can be applied to a subset of all the locations in a circuit, the noise strength can depend on the location, and it constrains the correlations only among faults of the same type (it does not limit the correlations of type-x faults with type-z faults). 

We can represent $\mathfrak{P}_{\rm 1-BP}$ as a directed depth-2 tree as in Fig.~\ref{fig:5}(b), where the level-1 noise locations are vertices at depth 1 below the root, and the level-0 noise locations are vertices at depth 2. Noise at the depth-1 locations is quasi-independent if we adopt the conventions explained in section \ref{sec:0-BPs}, and noise at the depth-2 locations is quasi-independent because the {\sc cnot} gates in the 1-BP preparation circuit are applied transversally. Furthermore, noise at a depth-1 location preceding one encoder is quasi-independent of noise at a depth-2 location following the other encoder---the 1-BP preparation circuit has been designed so that no fault can cause both an error in one output block and failure of the logical teleportation of the other output block.

We may say that vertex B on a directed tree is a ``descendant'' of vertex A if there is a directed path on the tree beginning at A and ending at B, and that a set of vertices is ``legitimate'' if no vertex in the set is a descendant of another vertex in the set. Then our observations are summarized by the following lemma:
\begin{lemm} \label{lemma:1}
Consider a legitimate subset $\mathcal{I}$ of the noisy locations in $\mathfrak{P}_{1-{\rm BP}}$. The noise in $\mathcal{I}$ is type-{\rm m} quasi-independent for ${\rm m = x, z}$. Conditioned on accepting the 1-BP, the noise strength is $\varepsilon_{\sigma_{\rm m}}(0,1|1)$ at depth-2 {\rm (}{\em i.e.}, level-0{\rm )} locations and $\varepsilon_{\sigma_{\rm m}}(1,1|1)$ at depth-1 {\rm (}{\em i.e.}, level-1{\rm )} locations, as specified in Eqs.~(\ref{eq:16}) and (\ref{eq:17}) below.
\end{lemm}

\noindent To estimate the conditional error probabilities $\varepsilon_{\sigma_{\rm m}}(0,1|1)$ and $\varepsilon_{\sigma_{\rm m}}(1,1|1)$, we will first estimate the joint probabilities for accepting the 1-BP together with errors at level 0 and level 1 respectively, then divide by an estimate of the acceptance probability.

Let $\varepsilon_{\sigma_{\rm m}}(0,1)$ denote the error strength disregarding whether the 1-BP is accepted, which is of course an upper bound on the joint probability for acceptance and a level-0 error. We can obtain an upper bound on $\varepsilon_{\sigma_{\rm m}}(0,1)$ by enumerating the locations in the circuit shown in Fig.~\ref{fig:2}(c) where type-m faults can cause type-m errors on qubits in the output blocks (we need only consider type-m faults since type-m errors preceding a {\sc cnot} gate are equivalent to type-m errors following the gate). Following the error propagation through the preparation circuit, we find
\begin{equation}
\label{eq:11}
\varepsilon_{\sigma_{\rm m}}(0,1) \leq \varepsilon + c_{1{\rm m}} \varepsilon_{\sigma_{\rm m}}(0,0) \; ,
\end{equation}
\noindent where $c_{1{\rm x}} = 1$ and $c_{1{\rm z}}=2$. The first term bounds the probability of a fault in a {\sc cnot} gate in the last step in the preparation circuit, and the second term bounds the probability that an error in a 0-BP propagates to an output qubit. Since $\sigma_{\rm x}$ errors do not propagate from the target to the control qubit of a {\sc cnot} gate while $\sigma_{\rm z}$ errors do, a $\sigma_{\rm x}$ error from just one 0-BP and $\sigma_{\rm z}$ errors from two 0-BPs may propagate to a given output qubit.   

Level-1 noise arises from errors in the logical teleportation of encoded blocks. Therefore, we obtain an upper bound on  $\varepsilon_{\sigma_{\rm m}}(1,1)$ by estimating the probability of an error in the  decoded measurement of a $C_4$ block. As described in section \ref{sec:decoding}, we propagate type-m errors due to faults in the circuit forward to the measurements, obtaining a quasi-independent description of the errors in the measured block; thus the probability of a logical error can be estimated using the function {\tt DECODE}. 

Let $b_{\sigma_{\rm m}}(0,1)$ denote the noise strength for the measured bits in the block. Analyzing the error propagation in the circuit shown in Fig.~\ref{fig:2}(c), we find
\begin{equation}
\label{eq:13}
b_{\sigma_{\rm m}}(0,1) \leq 4\, \varepsilon + c_{2{\rm m}} \varepsilon_{\sigma_{\rm m}}(0,0) \; ,
\end{equation}
\noindent where $c_{2{\rm x}} = 4$ and $c_{2{\rm z}}=3$. The first term bounds the probability of a fault in one of the three {\sc cnot} gates that act on each measured qubit or a fault in the measurement itself. The second term bounds the probability of an error in the qubit measurement caused by an error in a 0-BP---$\sigma_{\rm x}$ errors from at most four 0-BPs and $\sigma_{\rm z}$ errors from at most three 0-BPs can propagate to each measured qubit. 

The function {\tt DECODE} provides an upper bound on the probability that the measurement of a level-1 block is flagged, which we denote by $f^b_{\sigma_{\rm m}}(i{=}1,j{=}1)$. The superscript $b$ indicates that this is the flagging probability for block teleportation; later will use $f^s$ to denote the flagging probability for subblock teleportation (which has a different circuit). We let $b_{\sigma_{\rm m}}(i{=}1,j{=}1{\wedge}f)$ denote the joint probability that the block is flagged {\em and} the decoded measurement has a logical error, while $b_{\sigma_{\rm m}}(i{=}1,j{=}1{\wedge}{\neg f})$ denotes the joint probability that the block is {\em not} flagged {\em and} the decoded measurement has a logical error. Upper bounds on these quantities are computed using the function {\tt DECODE}, according to
\begin{widetext}
\begin{equation}
\label{eq:level-1-decode}
{\tt DECODE}: \{ 0, 0, b_{\sigma_{\rm m}}(0,1)\} \mapsto \{ f^b_{\sigma_{\rm m}}(1,1), b_{\sigma_{\rm m}}(1,1{\wedge}f), b_{\sigma_{\rm m}}(1,1{\wedge}{\neg f}) \}~.
\end{equation}
\end{widetext}
The first two arguments of {\tt DECODE} vanish because no level-0 bits are flagged.

If a flag is raised in the decoding of {\em any} block in the preparation circuit, we abort and the preparation of the 1-BP is restarted. Thus 
\begin{equation}
\label{eq:14b}
\varepsilon_{\sigma_{\rm m}} (1,1) \equiv b_{\sigma_{\rm m}} (1,1{\wedge}\neg f) \; 
\end{equation} 
is certainly an upper bound on the joint probability of acceptance together with a level-1 error, because the 1-BP will be accepted only if the level-1 measurement is unflagged.
\noindent For a flag to be raised inside the preparation circuit at least one fault must occur, so the probability of accepting a 1-BP is  
\begin{equation}
\label{eq:15}
p(j{=}1) \geq (1-\varepsilon )^{N} \; , 
\end{equation}
\noindent where $N=72$ is the number of elementary operations in the preparation circuit---there are 12 0-BPs, each of which is constructed by using 3 elementary operations, $5\times 4$ {\sc cnot} gates, and $4\times 4$ measurements. Therefore, we obtain an upper bound on the noise strength for the locations in $\mathfrak{P}_{\rm 1-BP}$, {\em conditioned} on the acceptance of the 1-BP: 
\begin{equation}
\label{eq:16}
\varepsilon_{\sigma_{\rm m}} (0,1 | 1) \leq  {\varepsilon_{\sigma_{\rm m}} (0,1) \over (1-\varepsilon )^N }
\end{equation} 
\noindent for the depth-2 noisy locations, with $\varepsilon_{\sigma_{\rm m}} (0,1)$ as in Eq.~(\ref{eq:11}), and 
\begin{equation}
\label{eq:17}
\varepsilon_{\sigma_{\rm m}} (1,1 | 1) \leq  {\varepsilon_{\sigma_{\rm m}} (1,1) \over (1-\varepsilon )^N }
\end{equation} 
\noindent for the depth-1 noisy locations, with $\varepsilon_{\sigma_{\rm m}} (1,1)$ as in Eq.~(\ref{eq:14b}). 

%---------------------------------------------------%
\subsection{Noise in $j$-BPs}
\label{sec:j-BPs}

\begin{figure}[t]
\begin{center}
\begin{tabular}{c}
\put(-3,0){\includegraphics[width=9cm,keepaspectratio]{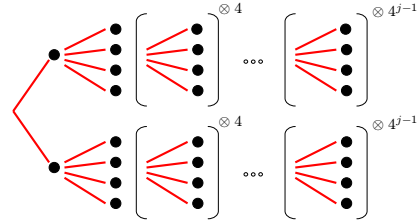}}
\vspace{-4cm}
\end{tabular}
\end{center}
\caption{\label{fig:6} The effective circuit $\mathfrak{P}_{j{-}{\rm BP}}$ that produces the same output as the $j$-BP preparation circuit in Fig.~\ref{fig:3}. The circuit is a depth-$(j{+}1)$ directed tree, with noisy locations at the vertices, and the noise is quasi-independent for any legitimate set of vertices. Conditioned on accepting the $j$-BP, the noise strength for errors of type m is  $\varepsilon_{\sigma_{\rm m}}(i,j|j)$ at depth-$(j{+}1{-}i)$ in the tree.
         }
\end{figure}
\begin{figure}[t]
\begin{center} 
\begin{tabular}{c}
\put(-4,0){\includegraphics[width=9cm,keepaspectratio]{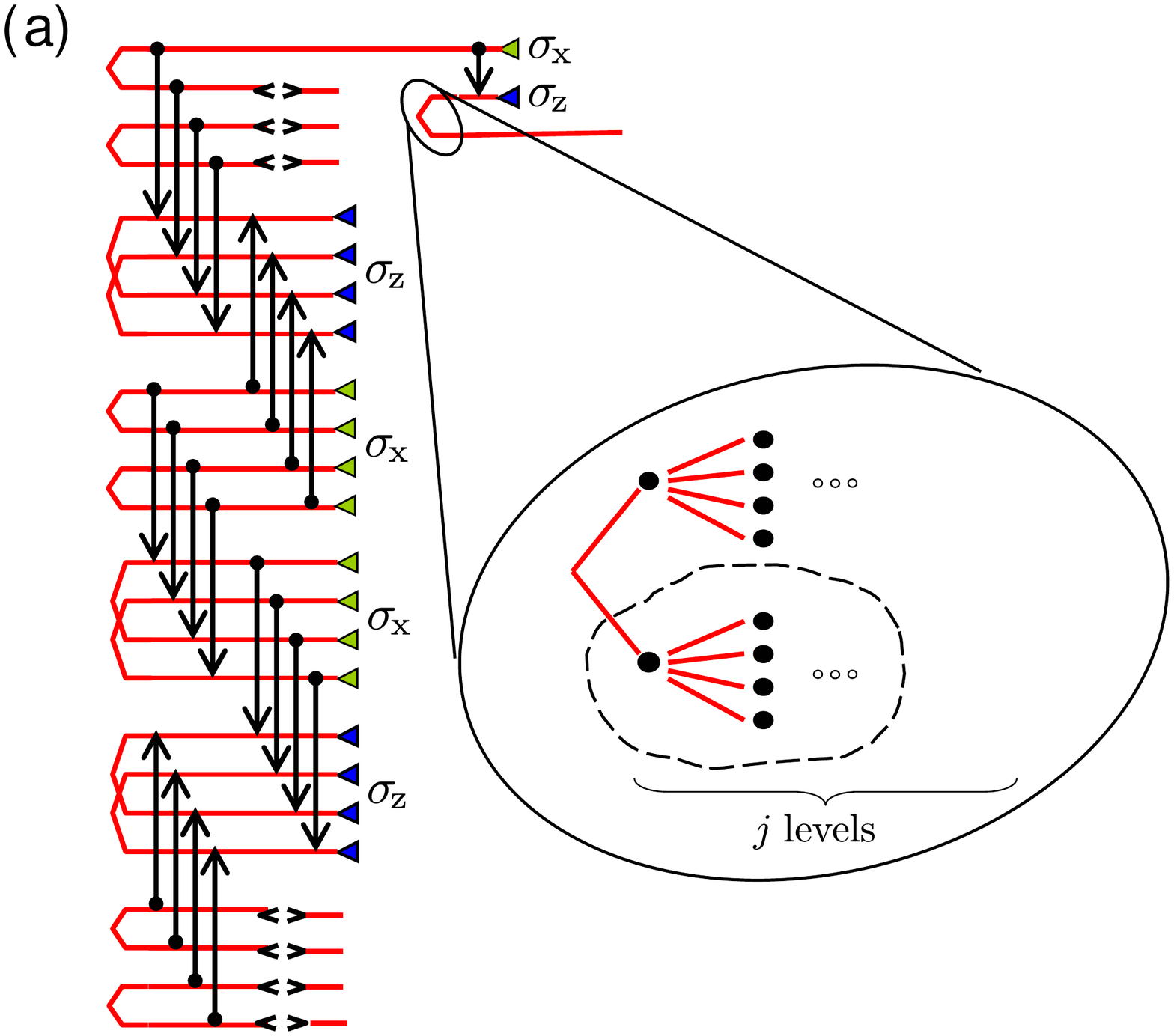}}
\put(-2.5,-6.5){\includegraphics[width=9cm,keepaspectratio]{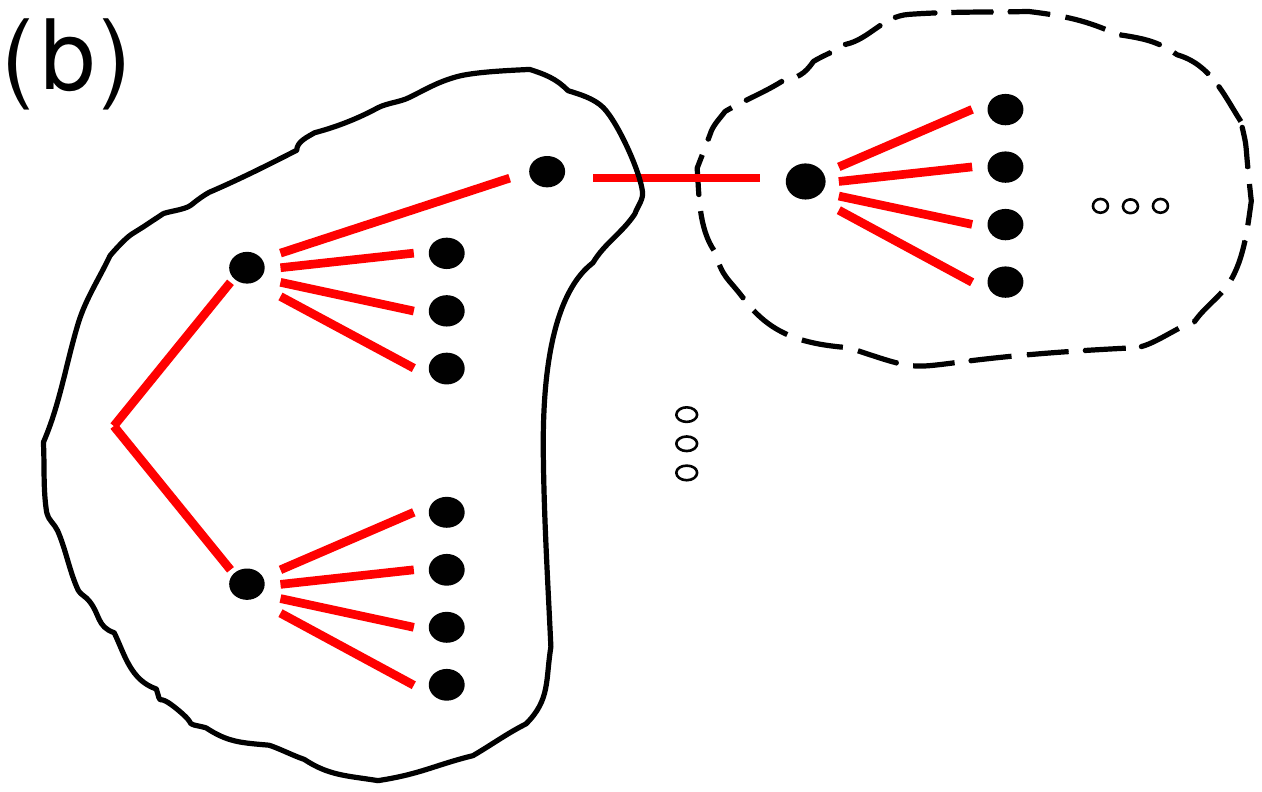}}
\vspace{-4cm}
\end{tabular}
\end{center}
\caption{\label{fig:7} (a) The $j$-BP preparation circuit, where each input state is a $(j{-}1)$-BP, each {\sc cnot} gate denotes a transversal {\sc cnot} acting on two $(C_4)^{\triangleright \, (j{-}1)}$ blocks, and each measurement denotes a transversal measurement of a $(C_4)^{\triangleright \, (j{-}1)}$ block. The magnified drawing shows the effective circuit $\mathfrak{P}_{(j{-}1){-}{\rm BP}}$, a depth-$j$ directed tree, that produces the same output as the $(j{-}1)$-BP preparation circuit; this $(j{-}1)$-BP is used to teleport a $(C_4)^{\triangleright \, j{-}1}$ subblock. (b) Construction of the effective circuit $\mathfrak{P}_{j{-}{\rm BP}}$ that produces the same output as the $j$-BP preparation circuit. Noisy locations at vertices of depth 3 or more are inherited from the $(j{-}1)$-BPs used in the subblock teleportations. The noise at depth 2 comes from two contributions which are combined together: the noise in the input $(j{-}1)$-BP (shown inside the dashed curve) and the noise due to logical errors in the subblock teleportation (shown inside the solid curve). The noise at depth 1 comes from logical errors in the block teleportation (also shown inside the solid curve). 
         }
\end{figure}

The preparation circuit of a $j$-BP with $j\ge 2$, shown in Fig.~\ref{fig:3}, includes subblock teleportations as well as block teleportations; otherwise it is similar to the preparation circuit of a 1-BP, but where the input Bell pairs are $(j{-}1)$-BPs instead of 0-BPs. As for the 1-BP, we obtain a hierarchical description of the noise in the $j$-BP preparation circuit by constructing the equivalent circuit $\mathfrak{P}_{j{-}{\rm BP}}$, shown schematically in Fig.~\ref{fig:6}. This equivalent circuit has the structure of a depth-$(j{+}1)$ directed tree, with level-$j$ noise at depth 1, level-$(j{-}1)$ noise at depth 2, {\em etc}., down to level-0 noise at depth $j{+}1$. Since the actual circuit in Fig.~\ref{fig:3} builds the $j$-BP starting with input $(j{-}1)$-BPs, we can characterize the noise in $\mathfrak{P}_{j{-}{\rm BP}}$ recursively. 

Our goal is to prove the following lemma:
\begin{lemm} \label{lemma:2}
Consider a legitimate subset $\mathcal{I}$ of the noisy locations in $\mathfrak{P}_{j{-}{\rm BP}}$. The noise in $\mathcal{I}$ is type-{\rm m} quasi-independent for ${\rm m = x, z}$. Conditioned on accepting the $j$-BP, the noise strength is $\varepsilon_{\sigma_{\rm m}}(i,j|j)$ at depth $j{+}1{-}i$ {\rm (}{\em i.e.}, level $i${\rm )} locations for $i=0,1,2,\dots, j$, as specified below.
\end{lemm}
\noindent We will estimate the conditional error probabilities using the identity
\begin{equation}
\label{eq:identity}
\varepsilon_{\sigma_{\rm m}}(i,j|j) = { \varepsilon_{\sigma_{\rm m}}(i,j|j{-}1) \over p(j|j{-}1)  } \; ;
\end{equation}
\noindent here, $\varepsilon_{\sigma_{\rm m}}(i,j|j{-}1)$ is the probability of a level-$i$ error in the output $j$-BP conditioned on only the acceptance of the input $(j{-}1)$-BPs, and  $p(j|j{-}1)$ is the probability of accepting the $j$-BP conditioned on the acceptance of the input $(j{-}1)$-BPs.

To construct the effective circuit $\mathfrak{P}_{j{-}{\rm BP}}$ for the preparation of the $j$-BP, we first replace each input $(j{-}1)$-BP in the $j$-BP preparation circuit by its effective circuit $\mathfrak{P}_{(j{-}1){-}{\rm BP}}$, as shown in Fig.~\ref{fig:7}(a). Because all of the output subblocks of the $j$-BP are teleported using $(j{-}1)$-BPs, the noise at level $i=0,1,2,\dots j{-}2$ in the $j$-BP is inherited directly from the noise in the $(j{-}1)$-BPs; hence for $0\le i\le j{-}2$ we have 
\begin{equation}
\label{eq:before-contraction}
\varepsilon_{\sigma_{\rm m}}(i,j|j{-}1) = \varepsilon_{\sigma_{\rm m}}(i,j{-}1|j{-}1)\; .
\end{equation}
However, the depth 2 ({\em i.e.}, level-$(j{-}1)$) noise in $\mathfrak{P}_{j{-}{\rm BP}}$ arises from two sources that must be combined together as shown in Fig.~\ref{fig:7}(b): errors occur both in the input $(j{-}1)$-BPs and in the teleportation of the $(C_4)^{\triangleright \, (j{-}1)}$ subblocks. Finally, the depth 1 ({\em i.e.}, level-$j$) noise in $\mathfrak{P}_{j{-}{\rm BP}}$ arises from errors in the teleportation of the $(C_4)^{\triangleright \, j}$ blocks.

First, we consider the block teleportation. The level-0 errors in the $(j{-}1)$-BPs and the level-0 errors due to faults in the {\sc cnot} gates can be propagated forward to the qubit measurements in the block teleportation. We find that errors of type m in the measurement outcomes (which are unflagged) have strength
%
%\begin{widetext}
\begin{equation}
b_{\sigma_{\rm m}}(0,j{\wedge}\neg f|j{-}1) \leq 4\, \varepsilon + c_{2{\rm m}} \varepsilon_{\sigma_{\rm m}}(0,j{-}1|j{-}1) \; ,
\vspace{.2cm}
\end{equation}
%\end{widetext}
%
\noindent where $c_{2{\rm x}} = 4$ and $c_{2{\rm z}}=3$ as in Eq.~(\ref{eq:13}). Since these errors are quasi-independent, we use the function {\tt DECODE} to find the error strength for the decoded $C_4$ block as in Eq.~(\ref{eq:level-1-decode}):
\begin{widetext}
\begin{equation}
\label{eq:level-1-decode-j}
{\tt DECODE}: \{0,0, b_{\sigma_{\rm m}}(0,j{\wedge}\neg f|j{-}1) \} \mapsto 
\{ f^b_{\sigma_{\rm m}}(1,j|j{-}1), \tilde{b}_{\sigma_{\rm m}}(1,j{\wedge}f | j{-}1), \tilde{b}_{\sigma_{\rm m}}(1,j{\wedge}{\neg f} | j{-}1) \}~,
\end{equation}
\end{widetext}
\noindent where all the probabilities are conditioned on the acceptance of the input $(j{-}1)$-BPs; here, $f^b_{\sigma_{\rm m}}(1,j|j{-}1)$ is an upper bound on the probability of flagging a $C_4$ subblock, $\tilde{b}_{\sigma_{\rm m}}(1,j{\wedge}f | j{-}1)$ is the joint probability that the subblock is flagged and there is a decoding error, and $\tilde{b}_{\sigma_{\rm m}}(1,j{\wedge}{\neg f} | j{-}1)$ is the joint probability that the subblock is not flagged and there is a decoding error. (We use $f^b$ to denote the flagging probability for a block teleportation; later we will use $f^s$ to denote the flagging probability for a subblock teleportation.)

But the measurement result for the $C_4$ block could be erroneous for either one of two reasons. The level-0 errors, propagated forward to the qubit measurements, could cause an error in the decoding of the $C_4$ subblock, and we have estimated the probability of such a decoding failure using the function {\tt DECODE}. Alternatively, a level-1 error in an input $(j{-}1)$-BP could propagate forward to a measured $C_4$ subblock, causing a logical error. We bound the total probability of a logical error in the decoded $C_4$ subblock by adding together the probabilities for these two sources of error, finding 
%
%\begin{widetext}
\begin{equation}
\begin{array}{rcl}
b_{\sigma_{\rm m}}(1,j{\wedge}g|j{-}1) & \leq & \tilde{b}_{\sigma_{\rm m}}(1,j{\wedge}g|j{-}1) \\
                                      &      & + c_{2{\rm m}} \varepsilon_{\sigma_{\rm m}}(1,j{-}1|j{-}1) \; ,
\end{array}
%\vspace{.2cm}
\end{equation}
%\end{widetext}
%
\noindent where $g{\in}\{f, \neg f \}$ designates whether the $C_4$ subblock is flagged or unflagged. Note that we use $\tilde{b}_{\sigma_{\rm m}}$ to denote the probability of a decoding error, and $b_{\sigma_{\rm m}}$ to denote the total probability of error, including the contribution from level-1 errors in the input $(j{-}1)$-BPs that propagate forward.

Now this procedure can be repeated. At level $i$, for $2\le i\le j{-}1$, we apply {\tt DECODE} to the level-$(i{-}1)$ error and flagging probabilities to estimate the probability $\tilde{b}_{\sigma_{\rm m}}(i,j{\wedge}g|j{-}1)$ of a level-$i$ decoding error; then we add the estimated probability for a level-$i$ error in an input $(j{-}1)$-BP to propagate to the measured level-$i$ subblock, obtaining our estimate of the total probability $b_{\sigma_{\rm m}}(i,j{\wedge}g|j{-}1)$ of error for the measured $(C_4)^{\triangleright \, i}$ subblock:
%
%\begin{widetext}
\begin{equation}
\begin{array}{rcl}
b_{\sigma_{\rm m}}(i,j{\wedge}g|j{-}1) & \leq & \tilde{b}_{\sigma_{\rm m}}(i,j{\wedge}g|j{-}1) \\
                                       &      &+ c_{2{\rm m}} \varepsilon_{\sigma_{\rm m}}(i,j{-}1|j{-}1) \; .
\end{array}
\end{equation}
%\end{widetext}
%
\noindent At the top ($i=j$) level, there is no contribution from errors in the input $(j{-}1)$-BPs; furthermore, the preparation of the $j$-BP is aborted if a flag is raised in the final decoding step, so we need only consider the case where there is no flag, finding 
\begin{equation}
\varepsilon_{\sigma_{\rm m}} (j,j|j{-}1) \le \tilde{b}_{\sigma_{\rm m}}(j,j{\wedge}\neg f|j{-}1) \;. 
\end{equation} 
\noindent The recursion relations from which we derive our upper bound on $\varepsilon_{\sigma_{\rm m}} (j,j|j{-}1)$ are summarized in Table~\ref{table:1}.

%--------------------------------------------------------------------%

\begin{table*}
\begin{center}
\begin{tabular}{l}
\hline \hline

\noindent {\sc step 1: } \\

\begin{minipage}{16cm} \begin{flushleft}

Measurements of level-0 qubits are unflagged; therefore, $f^b_{\sigma_{\rm m}}(0,j|j{-}1) = 0 = b_{\sigma_{\rm m}}(0,j{\wedge}f|j{-}1)$. 

\vspace{.1cm}
Analyzing the error propagation in the $j$-BP preparation circuit shown in Fig.~\ref{fig:3}, we estimate the error strength for the level-0 measurements in the block teleportation:
\[
b_{\sigma_{\rm m}}(0,j{\wedge}\neg f|j{-}1) \leq 4\, \varepsilon + c_{2{\rm m}} \varepsilon_{\sigma_{\rm m}}(0,j{-}1|j{-}1) \; ,
\]
\noindent where $c_{2{\rm x}} = 4$ and $c_{2{\rm z}}=3$. \vspace{.1cm}

\end{flushleft} \end{minipage} \\ \hline

\begin{minipage}{16cm} \begin{flushleft}

\vspace{.1cm}
\noindent {\sc step 2:} 

\noindent $(a)$ Set $i=0$.

\vspace{.1cm}
\noindent $(b)$ Using the function {\tt DECODE}, we map the flagging and error probabilities at level $i$ to the flagging and decoding-error probabilities at level $i{+}1$:
\[
\begin{array}{rl}
{\tt DECODE} : & \{ \;\;\;\;\, f^b_{\sigma_{\rm m}}(i,j|j{-}1), \;\;\;\;\, b_{\sigma_{\rm m}}(i,j{\wedge}f|j{-}1), \;\;\;\;\, b_{\sigma_{\rm m}}(i,j{\wedge}\neg f|j{-}1)\}  \\
\mapsto & \{ f^b_{\sigma_{\rm m}}(i{+}1,j|j{-}1), \tilde{b}_{\sigma_{\rm m}}(i{+}1,j{\wedge}f | j{-}1), \tilde{b}_{\sigma_{\rm m}}(i{+}1,j{\wedge}{\neg f} | j{-}1) \} \; .
\end{array}
\]

\noindent $(c)$ We propagate the level-$(i{+}1)$ errors in the input $(j{-}1)$-BPs forward to the measured level-$(i{+}1)$ subblocks, and combine them with the level-$(i{+}1)$ decoding errors to estimate the total level-$(i{+}1)$ noise strength:
\[
b_{\sigma_{\rm m}}(i{+}1,j{\wedge}g|j{-}1) \leq \tilde{b}_{\sigma_{\rm m}}(i{+}1,j{\wedge}g|j{-}1) + c_{2{\rm m}} \varepsilon_{\sigma_{\rm m}}(i{+}1,j{-}1|j{-}1) \; ,
\]
\noindent where $g{\in}\{f, \neg f \}$ designates whether the level-$(i{+}1)$ subblock is flagged or unflagged. \vspace{.1cm}

\end{flushleft} \end{minipage} \\ \hline

\begin{minipage}{16cm} \begin{flushleft}

\vspace{.1cm}
\noindent {\sc step 3:} 

If $i{+}1<j{-}1$, increase $i$ by 1, go to step $2(b)$, and repeat. \vspace{.1cm}

\end{flushleft} \end{minipage} \\ \hline

\begin{minipage}{16cm} \begin{flushleft}

\vspace{.1cm}
\noindent {\sc step 4:} 

Apply step $2(b)$ for $i = j{-}1$. If the level-$j$ block is flagged, the preparation is aborted. Thus we consider only the case where there is no flag, obtaining the level-$j$ error strength in the $j$-BP: 
\[
\varepsilon_{\sigma_{\rm m}}(j,j|j{-}1) \equiv \tilde{b}_{\sigma_{\rm m}}(j,j{\wedge}\neg f|j{-}1) \; . \vspace{.2cm}
\]

\end{flushleft} \end{minipage} \\ \hline \hline

\end{tabular}
\end{center}
\vspace{-.4cm}
\caption{\label{table:1} Recursion equations for estimating the level-$j$ noise strength in a $j$-BP.}
\end{table*}

%----------------------------------------------------%

The recursion relations for deriving an upper bound on the probability of error in the subblock teleportation are derived by exactly the same logic, and are summarized in Table~\ref{table:2}. The details are different because the circuit for subblock teleportation differs from the circuit for block teleportation. For type-m errors in the measured level-0 qubits the error strength is
%
%\begin{widetext}
\begin{equation}
\begin{array}{c}
s_{\sigma_{\rm m}}(0,j{\wedge}\neg f|j{-}1) \leq 3\, \varepsilon{+}(1{+}c_{1{\rm m}}) \varepsilon_{\sigma_{\rm m}}(0,j{-}1|j{-}1) \, ,
\end{array}
\vspace{.2cm}
\end{equation}
%\end{widetext}

\noindent where $c_{1{\rm x}} = 1$ and $c_{1{\rm z}}=2$ as in Eq.~(\ref{eq:11}). The first term bounds the probability of a fault in one of the two {\sc cnot} gates that act on each measured qubit or a fault in the measurement itself. The second term bounds the probability of a level-0 error in a $(j{-}1)$-BP that propagates forward to the qubit measurement---$\sigma_{\rm x}$ errors from two $(j{-}1)$-BPs and $\sigma_{\rm z}$ errors from three $(j{-}1)$-BPs can propagate to a given measured qubit. Of course, the level-0 bits are all unflagged. 

At each level $i$, where $1\le i\le j{-}1$, we apply the function {\tt DECODE} to the flagging and error probabilities at level $i{-}1$ to estimate the level-$i$ flagging probability $f^s_{\sigma_{\rm m}}(i,j|j{-}1)$ and the decoding-error probabilities $\tilde{s}_{\sigma_{\rm m}}(i,j{\wedge}g | j{-}1)$, where $g{\in}\{f, \neg f \}$. Then we add the estimated probability for a level-$i$ error in an input $(j{-}1)$-BP to propagate to the measured level-$i$ subblock, obtaining our estimate of the total probability $s_{\sigma_{\rm m}}(i,j{\wedge}g|j{-}1)$ of error for the measured $(C_4)^{\triangleright \, i}$ subblock:
%
%\begin{widetext}
\begin{equation}
\begin{array}{rcl}
s_{\sigma_{\rm m}}(i,j{\wedge}g|j{-}1) & \leq & \tilde{s}_{\sigma_{\rm m}}(i,j{\wedge}g|j{-}1) \\
                                       &      & + (1{+}c_{1{\rm m}}) \varepsilon_{\sigma_{\rm m}}(i,j{-}1|j{-}1) \; .
\end{array}    
\vspace{.2cm}                                   
\end{equation}
%\end{widetext}
%
Repeating this procedure until reaching the top ($i=j{-}1$) level, we estimate the probability of a type-m logical error for the top-level measurement in the teleportation of the $(C_4)^{\triangleright \, (j{-}1)}$ subblock. Since the preparation of the $j$-BP is aborted if a flag is raised at the top-level in any subblock teleportation, we need only consider the noise strength $s_{\sigma_{\rm m}}(j{-}1,j{\wedge}\neg f|j{-}1)$ for the case where there is no flag.

Finally, in a level-$(j{-}1)$ subblock of the output $j$-BP, there are two possible sources of error, as indicated in Fig.~\ref{fig:7}(b). For each subblock teleportation, there is an unmeasured output $(C_4)^{\triangleright \, (j{-}1)}$ subblock and a measured $(C_4)^{\triangleright \, (j{-}1)}$ subblock. So far we have estimated only the error probability $s_{\sigma_{\rm m}}(j{-}1,j{\wedge}\neg f|j{-}1)$ for the measurement; to estimate the total level-$(j{-}1)$ error strength $\varepsilon_{\sigma_{\rm m}} (j{-}1,j|j{-}1)$ for a level-$(j{-}1)$ subblock of the $j$-BP, we should add the estimate of the measurement-error probability to our estimate $\varepsilon_{\sigma_{\rm m}}(j{-}1,j{-}1|j{-}1)$ of the top-level error strength in the unmeasured output block of the $(j{-}1)$-BP, finding:
%
%\begin{widetext}
\begin{equation}
\begin{array}{rcl}
\varepsilon_{\sigma_{\rm m}} (j{-}1,j|j{-}1) & \leq & s_{\sigma_{\rm m}}(j,j{\wedge}\neg f|j{-}1) \\
                                             &      & + \varepsilon_{\sigma_{\rm m}}(j{-}1,j{-}1|j{-}1) \;.
\end{array}
\end{equation}
%\end{widetext}

%----------------------------------------------------%

\begin{table*}
\begin{center}
\begin{tabular}{l}
\hline \hline

\noindent {\sc step 1:} \\

\begin{minipage}{16cm} \begin{flushleft}

Measurements of level-0 qubits are unflagged; therefore,  $f^s_{\sigma_{\rm m}}(0,j|j{-}1) = 0 =s_{\sigma_{\rm m}}(0,j{\wedge}f|j{-}1)$.

\vspace{.1cm}
Analyzing the error propagation in the $j$-BP preparation circuit shown in Fig.~\ref{fig:3}, we estimate the error strength for the level-0 measurements in the subblock teleportation:
\[
s_{\sigma_{\rm m}}(0,j{\wedge}\neg f|j{-}1) \leq 3\, \varepsilon + \left(1+c_{1{\rm m}} \right) \varepsilon_{\sigma_{\rm m}}(0,j{-}1|j{-}1) \; , 
\] 
\noindent where $c_{1{\rm x}}=1$ and $c_{1{\rm z}}=2$. \vspace{.1cm}

\end{flushleft} \end{minipage} \\ \hline

\begin{minipage}{16cm} \begin{flushleft}

\vspace{.1cm}
\noindent {\sc step 2:} 

\noindent $(a)$ Set $i=0$.

\vspace{.1cm}
\noindent $(b)$ Using the function {\tt DECODE}, we map the flagging and error probabilities at level $i$ to the flagging and decoding-error probabilities at level $i{+}1$:
\[
\begin{array}{rl}
{\tt DECODE} : & \{ \;\;\;\;\, f^s_{\sigma_{\rm m}}(i,j|j{-}1), \;\;\;\;\, s_{\sigma_{\rm m}}(i,j{\wedge}f|j{-}1), \;\;\;\;\, s_{\sigma_{\rm m}}(i,j{\wedge}\neg f|j{-}1)\}  \\
\mapsto & \{ f^s_{\sigma_{\rm m}}(i{+}1,j|j{-}1), \tilde{s}_{\sigma_{\rm m}}(i{+}1,j{\wedge}f | j{-}1), \tilde{s}_{\sigma_{\rm m}}(i{+}1,j{\wedge}{\neg f} | j{-}1) \} \; .
\end{array}
\]

\noindent $(c)$ We propagate the level-$(i{+}1)$ errors in the input $(j{-}1)$-BPs forward to the measured level-$(i{+}1)$ subblocks, and combine them with the level-$(i{+}1)$ decoding errors to estimate the total level-$(i{+}1)$ noise strength:
\[
s_{\sigma_{\rm m}}(i{+}1,j{\wedge}g|j{-}1) \leq \tilde{s}_{\sigma_{\rm m}}(i{+}1,j{\wedge}g|j{-}1) + \left(1+c_{1{\rm m}} \right)  \varepsilon_{\sigma_{\rm m}}(i{+}1,j{-}1|j{-}1) \; ,
\]
\noindent where $g{\in}\{f, \neg f \}$ designates whether the level-$(i{+}1)$ subblock is flagged or unflagged. \vspace{.1cm}

\end{flushleft} \end{minipage} \\ \hline

\begin{minipage}{16cm} \begin{flushleft}

\vspace{.1cm}
\noindent {\sc step 3:} 

If $i{+}1<j{-}1$, increase $i$ by 1, go to step $2(b)$, and repeat. \vspace{.1cm}

\end{flushleft} \end{minipage} \\ \hline

\begin{minipage}{16cm} \begin{flushleft}

\vspace{.1cm}
\noindent {\sc step 4:} 

If the level-$(j{-}1)$ block is flagged, the preparation is aborted. Thus we consider only the case where there is no flag, obtaining the error strength $s_{\sigma_{\rm m}}(j{-}1,j{\wedge}\neg f|j{-}1)$ for the level-$(j{-}1)$ measurement in the subblock teleportation. Combining with the top-level error strength $\varepsilon_{\sigma_{\rm m}}(j{-}1,j{-}1|j{-}1)$ for an output block of the $(j{-}1)$-BP, we obtain our estimate of the total level-$(j{-}1)$ error strength in the $j$-BP:
\[
\varepsilon_{\sigma_{\rm m}}(j{-}1,j|j{-}1) \leq s_{\sigma_{\rm m}}(j{-}1,j{\wedge}\neg f|j{-}1) + \varepsilon_{\sigma_{\rm m}}(j{-}1,j{-}1|j{-}1) \; . \vspace{.2cm}
\]

\end{flushleft} \end{minipage} \\ \hline \hline

\end{tabular}
\end{center}
\vspace{-.4cm}
\caption{\label{table:2} Recursion equations for estimating the level-$(j{-}1)$ noise strength in a $j$-BP.}
\end{table*}

%-------------------------------------------------------------------------%

Until now, our estimated error strengths $\varepsilon_{\sigma_{\rm m}} (i,j|j{-}1)$ ($i=0,1,2,\dots, j$) have been conditioned on only the acceptance of the input $(j{-}1)$-BPs. To find estimates of the error strengths conditioned on the acceptance of the $j$-BP, we divide by an estimate of the acceptance probability $p(j|j{-}1)$ as in Eq.~(\ref{eq:identity}). We recall that a $j$-BP is accepted only if no flag is raised in the final decoding step in either of the two block teleportations or any of the eight subblock teleportations. Therefore, the probability $p(j|j{-}1)$ that a $j$-BP is accepted, conditioned on having accepted all input $(j{-}1)$-BPs, satisfies
\begin{widetext}
\begin{equation}
\label{eq:18}
p(j|j{-}1) \geq 1 - \sum\limits_{{\rm m}\in \{ {\rm x}, {\rm z} \} } \left( 2\, f^b_{\sigma_{\rm m}}(j,j|j{-}1) + 8\, f^s_{\sigma_{\rm m}}(j{-}1,j|j{-}1) \right) \; .
\end{equation}
\end{widetext}

%---------------------------------------------------%

\section{Level reduction for $\mathcal{G}_{\rm CSS}$ gadgets}
\label{sec:CSS-gadgets}

With an adequate hierarchical characterization of the errors in a $j$-BP now in hand, we can proceed to analyze the reliability of a circuit constructed from level-$j$ $\mathcal{G}_{\rm CSS}$ gadgets protected by the code $(C_4)^{\triangleright \, j}$. For the {\sc cnot} gadget, we consider the level-$j$ version of the circuit shown in Fig.~\ref{fig:1}, where each 1-BP is replaced by a $j$-BP, each {\sc cnot} gate denotes a transversal {\sc cnot} acting on two $(C_4)^{\triangleright \, (j{-}1)}$ blocks, and each measurement denotes a transversal measurement of a $(C_4)^{\triangleright \, (j{-}1)}$ block. We represent each $j$-BP preparation circuit by the equivalent circuit $\mathfrak{P}_{j{-}{\rm BP}}$, a depth-$(j{+}1)$ directed tree with noisy locations at each coding level. Then we propagate errors in the effective {\sc cnot} gadget (both the errors due to faults in the circuit and errors in the input $j$-BPs) forward to the measurements inside the block teleportations, to estimate the measurement failure probability. If all decoded level-$j$ measurements in the teleportations of the gadget's {\em output} blocks agree with the measurements in the ideal noiseless circuit, then the {\sc cnot} gadget simulates an ideal {\sc cnot} gate successfully; otherwise the gadget fails. From our estimate of the probability that a level-$j$ measurement fails, then, we obtain an estimate of the probability of failure $\varepsilon_{\rm css}(j)$ for a level-$j$ {\sc cnot} gadget.

We would like to regard $\varepsilon_{\rm css}(j)$ as the noise strength for an effective local stochastic noise model that characterizes the reliability of the level-$j$ simulated circuit. But our procedure for estimating the gadget failure probability still requires an important modification, to ensure that the probability of failure for a pair of consecutive gadgets is bounded above by $\big(\varepsilon_{\rm css}(j)\big)^2$. Errors in a $j$-BP that is used for the teleportation of one of the {\em input} blocks in a {\sc cnot} gadget can propagate to the measurements in teleportations of {\em both} the input and output blocks, causing correlated logical errors in two consecutive gadgets. We address this problem using the {\em truncation} method introduced in \cite{Aliferis05b}. For each fault path, we classify each gadget as ``good'' or ``bad'' starting at the back of the circuit, and moving toward the front step by step. A {\sc cnot} gadget that is followed by two good gadgets is declared bad if it contains a failed measurement in the teleportation of at least one output block; otherwise it is good. A {\sc cnot} gadget that is followed by one good gadget and one bad gadget is declared bad only if it contains a failed measurement in the teleportation of the output block that is an input to the following {\em good} gadget. A {\sc cnot} gadget that is followed by two bad gadgets is always declared good.

With these definitions, the sum of the probabilities of all fault paths such that $r$ specified level-$j$ gadgets are all bad is bounded above by $\big(\varepsilon_{\rm css}(j)\big)^r$. Furthermore, we may say that a good {\sc cnot} gadget simulates the ideal {\sc cnot} gate correctly. In the case of a good {\sc cnot} gadget followed by one bad gadget and one good gadget, an error in the teleportation of the output block that is an input block in the following bad gadget can be propagated forward into the bad gadget. Likewise, in the case of a good {\sc cnot} gadget followed by two bad gadgets, errors in the teleportations of the two output blocks can be propagated forward into the following bad gadgets.

Aside from {\sc cnot} gadgets, $\mathcal{G}_{\rm CSS}$ circuits also include ``contracted gadgets,'' in which state preparation steps are combined with following {\sc cnot} gates to form composite gadgets, or measurement steps are combined with preceding {\sc cnot} gates to form composite gadgets \cite{Aliferis07b}. These contracted gadgets are smaller and more reliable than {\sc cnot} gadgets; therefore, to estimate the effective noise strength in a simulated $\mathcal{G}_{\rm CSS}$ circuit, we may limit our attention to the analysis of the {\sc cnot} gadget.

Therefore we can prove the following lemma:

\begin{lemm} \label{lemma:3} 
Consider a level-$j$ simulation of a $\mathcal{G}_{\rm CSS}$ circuit, where elementary operations are subject to local stochastic noise with strength $\varepsilon$. There is an equivalent level-0 $\mathcal{G}_{\rm CSS}$ circuit, producing the same output, which is subject to local stochastic noise with strength $\varepsilon_{\rm css}(j)$, as specified in Table~\ref{table:3}.
\end{lemm}

%---------------------------------------------------------%

\begin{table*}
\begin{center}
\begin{tabular}{l}
\hline \hline

\noindent {\sc step 1:} \\

\begin{minipage}{16cm} \begin{flushleft}

Measurements of level-0 qubits are unflagged; therefore, $f^r_{\sigma_{\rm m}}(0,j|j) = 0 = r_{\sigma_{\rm m}}(0,j{\wedge}f|j)$.

\vspace{.1cm}
Analyzing the error propagation in the {\sc cnot} gadget shown in Fig.~\ref{fig:1}, we estimate the error strength for level-0 measurements:
\[
r_{\sigma_{\rm m}}(0,j{\wedge}\neg f|j) \leq 3\, \varepsilon + c_{3{\rm rm}} \varepsilon_{\sigma_{\rm m}}(0,j|j) \; , 
\]
\noindent where $c_{3{\rm cx}}=c_{3{\rm tz}}=2$ and $c_{3{\rm cz}} = c_{3{\rm tx}} = 3$. \vspace{.1cm}

\end{flushleft} \end{minipage} \\ \hline

\begin{minipage}{16cm} \begin{flushleft}

\vspace{.1cm}
\noindent {\sc step 2:} 

\noindent $(a)$ Set $i=0$.

\vspace{.1cm}
\noindent $(b)$ Using the function decode, we map the flagging and error probabilities at level $i$ to the flagging and decoding-error probabilities at level $i{+}1$:
\[
\begin{array}{rl}
{\tt DECODE} : & \{ \;\;\;\;\, f^r_{\sigma_{\rm m}}(i,j|j), \;\;\;\;\, r_{\sigma_{\rm m}}(i,j{\wedge}f|j), \;\;\;\;\, r_{\sigma_{\rm m}}(i,j{\wedge}\neg f|j)\}  \\
\mapsto & \{ f^r_{\sigma_{\rm m}}(i{+}1,j|j), \tilde{r}_{\sigma_{\rm m}}(i{+}1,j{\wedge}f | j), \tilde{r}_{\sigma_{\rm m}}(i{+}1,j{\wedge}{\neg f} | j) \} \; .
\end{array}
\]

\noindent $(c)$ We propagate the level-$(i{+}1)$ errors in the input $j$-BPs forward to the measured level-$(i{+}1)$ subblocks, and combine them with the level-$(i{+}1)$ decoding errors to estimate the total level-$(i{+}1)$ noise strength:
\[
r_{\sigma_{\rm m}}(i{+}1,j{\wedge}g|j) \leq \tilde{r}_{\sigma_{\rm m}}(i{+}1,j{\wedge}g|j) + c_{3{\rm rm}} \varepsilon_{\sigma_{\rm m}}(i{+}1,j|j) \; ,
\]
\noindent where $g{\in}\{f, \neg f \}$ designates whether the level-$(i{+}1)$ subblock is flagged or unflagged. \vspace{.1cm}

\end{flushleft} \end{minipage} \\ \hline

\begin{minipage}{16cm} \begin{flushleft}

\vspace{.1cm}
\noindent {\sc step 3:} 

If $i{+}1<j$, increase $i$ by 1, go to step $2(b)$, and repeat. \vspace{.1cm}

\end{flushleft} \end{minipage} \\ \hline

\begin{minipage}{16cm} \begin{flushleft}

\vspace{.1cm}
\noindent {\sc step 4:} 

Measurements are accepted whether or not a flag is raised in the final decoding step. Therefore, the noise strength for a level-j measurement is
\[
r_{\sigma_{\rm m}}(j,j|j) = r_{\sigma_{\rm m}}(j,j{\wedge}f|j) + r_{\sigma_{\rm m}}(j,j{\wedge}\neg f|j) \; .
\]
\noindent A $\mathcal{G}_{\rm CSS}$ gadget fails if there is a measurement error in the teleportation of at least one of its output blocks; therefore,
\[
\varepsilon^{\;\;}_{\rm css}(j) \leq \sum\limits_{ r\in \{ {\rm c}, {\rm t} \}} \sum\limits_{{\rm m}\in\{ {\rm x}, {\rm z} \} } r_{\sigma_{\rm m}}(j,j|j) \; . \vspace{.2cm}
\]

\end{flushleft} \end{minipage} \\ \hline \hline

\end{tabular}
\end{center}
\vspace{-.4cm}
\caption{\label{table:3} Recursion equations for estimating the effective noise strength of a level-$j$ $\mathcal{G}_{\rm CSS}$ circuit.}
\end{table*}
%-----------------------------------------%

We estimate the effective noise strength $\varepsilon_{\rm css}(j)$ of the level-$j$ $\mathcal{G}_{\rm CSS}$ circuit using a recursive procedure, summarized in Table~\ref{table:3}, that closely follows our method for analyzing the $j$-BP preparation. We represent the preparation of the four $j$-BPs in the {\sc cnot} gadget by effective depth-$(j{+}1)$ circuits $\mathfrak{P}_{j{-}{\rm BP}}$, and we propagate level-0 errors in the $j$-BPs and level-0 errors due to faults in the circuit forward to the qubit measurements in the teleportations of the outgoing $(C_4)^{\triangleright \, j}$ blocks; thus we obtain an upper bound on the noise strength for these (unflagged) measurements:
\begin{equation}
r_{\sigma_{\rm m}}(0,j{\wedge}\neg f|j) \leq 3\, \varepsilon + c_{3{\rm rm}} \varepsilon_{\sigma_{\rm m}}(0,j|j) \; , 
\end{equation}
\noindent where ${\rm r}\in \{{\rm c},{\rm t} \}$ designates whether the measured output block is the control or target block, and $c_{3{\rm cx}}=2$, $c_{3{\rm tz}}=2$, $c_{3{\rm cz}} = 3$, $c_{3{\rm tx}} = 3$. The first term bounds the probability of a fault in one of the two {\sc cnot} gates that act on each measured qubit or a fault in the measurement itself. The second term bounds the probability of a level-0 error in an input $j$-BP that propagates forward to the qubit measurement---$\sigma_{\rm x}$ errors in the input control block and $\sigma_{\rm z}$ errors in the input target block can propagate to both the output control and the output target block, while $\sigma_{\rm x}$ errors in the input target block and $\sigma_{\rm z}$ errors in the input control block do not propagate to the other output block. 

At each level $i$, where $1\le i\le j$, we apply the function {\tt DECODE} to the flagging and error probabilities at level $i{-}1$ to estimate the level-$i$ flagging probability $f^r_{\sigma_{\rm m}}(i,j|j)$ and the decoding-error probabilities $\tilde{r}_{\sigma_{\rm m}}(i,j{\wedge}g | j)$, where $g{\in}\{f, \neg f \}$; these probabilities are conditioned on accepting the input $j$-BPs. Then we add the estimated probability for a level-$i$ error in an input $j$-BP to propagate to the measured level-$i$ subblock, obtaining our estimate of the total probability $r_{\sigma_{\rm m}}(i,j{\wedge}g | j)$ of error for the measured $(C_4)^{\triangleright \, i}$ subblock:
%
%\begin{widetext}
\begin{equation}
\begin{array}{rcl}
r_{\sigma_{\rm m}}(i,j{\wedge}g|j) & \leq & \tilde{r}_{\sigma_{\rm m}}(i,j{\wedge}g|j) \\
                                   &      & + c_{3{\rm rm}} \varepsilon_{\sigma_{\rm m}}(i,j|j) \; .
\end{array}                                   
\end{equation}
%\end{widetext}
%
There is no postselection inside gadgets, so the top-level ($i=j$) measurement is accepted whether or not a flag is raised in the final decoding step. Therefore, the error probability for the measurement of a level-$j$ block is bounded above by
\begin{equation}
r_{\sigma_{\rm m}}(j,j|j) = r_{\sigma_{\rm m}}(j,j{\wedge}f|j) + r_{\sigma_{\rm m}}(j,j{\wedge}\neg f|j) \; .
\end{equation}

The {\sc cnot} gadget fails if there is an error in either the $\sigma_{\rm x}$ or $\sigma_{\rm z}$ measurement in the teleportation of either the control or the target block. Summing over these four measurements, our estimate of the failure probability for the level-$j$ {\sc cnot} gadget becomes:
\begin{equation}
\varepsilon^{\;\;}_{\rm css}(j) \leq \sum\limits_{ r\in \{ {\rm c}, {\rm t} \}} \sum\limits_{{\rm m}\in\{ {\rm x}, {\rm z} \} } r_{\sigma_{\rm m}}(j,j|j) \; .
\end{equation}
%

%---------------------------------------------------%

\section{The threshold for $\mathcal{G}_{\rm CSS}$ operations}
\label{sec:CSS-threshold}

\subsection{The $\mathcal{G}_{\rm CSS}$ threshold for local stochastic noise}

By analyzing the recursion equations, first for $j$-BPs and then for level-$j$ $\mathcal{G}_{\rm CSS}$ gadgets, we can obtain an upper bound on the effective noise strength $\varepsilon^{\;\;}_{\rm css}(j)$ in the level-$j$ simulation of a $\mathcal{G}_{\rm CSS}$ circuit, as a function of $j$ and the fundamental noise strength $\varepsilon$. Before we present these upper bounds, we discuss a few simple optimizations that help us improve our results.

First, we discuss an improvement in the analysis of noise in 1-BPs in section \ref{sec:1-BPs}. Errors in some of the input 0-BPs propagate not only to the output qubits of the 1-BP but also to the measurements. Since the 1-BP is accepted only if no errors are detected, we can exclude some fault paths and refine our estimate of the coefficient $c_1$ in Eq.~(\ref{eq:11}). For example, a $\sigma_{\rm x}$ error in a 0-BP in an unmeasured block will propagate to a $\sigma_{\rm z}$ measurement and be detected unless another $\sigma_{\rm x}$ error occurs due to a fault in another operation. By considering how errors propagate in the preparation circuit, we obtain the improved estimate
\begin{equation}
\label{eq:c1-improve}
c_1 = \begin{cases} \; 16\, \varepsilon_{\sigma_{\rm x}}(0,0) + 16\,\varepsilon \;, & \hspace{-.3cm} {\rm if} \; {\rm m}={\rm x} \\ \; 1 + 16\, \varepsilon_{\sigma_{\rm z}}(0,0) + 16\, \varepsilon \;, & \hspace{-.3cm} {\rm if} \; {\rm m}={\rm z} \end{cases} \; ;
\end{equation}
\noindent here, $16\, \varepsilon_{\sigma_{\rm m}}(0,0)$ is an upper bound on the probability that an error in a 0-BP propagates to cause a measurement error for at least one of the qubits in a given block;  there are at most 8 such 0-BPs, and for each 0-BP noise of strength  $\varepsilon_{\sigma_{\rm m}}(0,0)$ acts on each of its output qubits.  Also, $16\,\varepsilon$ is an upper bound on the probability that a fault in the circuit causes a measurement error for at least one of the qubits in a given block; faults in any of $3\times 4$ {\sc cnot} gates and 4 qubit measurements could cause the error.

\begin{figure}[t]
\begin{center} 
\includegraphics[width=8.3cm,keepaspectratio]{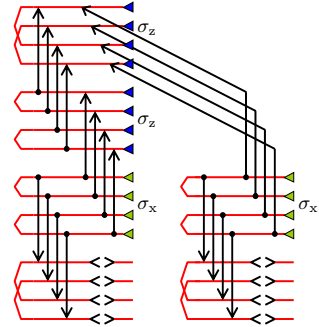}
\vspace{-2.5cm} 
\end{center}
\caption{\label{fig:10} An alternative preparation circuit for $j$-BPs. Compared to the circuit in Fig.~\ref{fig:3}, the direction of the {\sc cnot} gates is reversed so that type-x and type-z errors propagate differently than in Fig.~\ref{fig:3}.     
         }
\end{figure}

Secondly, we observe that the circuit in Fig.~\ref{fig:3} which we use for preparing $j$-BPs treats the two types of errors differently; this is reflected in the coefficients in our recursion equations where $c_{1{\rm x}}\not = c_{1{\rm z}}$ and $c_{2{\rm x}}\not = c_{2{\rm z}}$. To weaken the asymmetry between the different types of errors, we modify our recursive preparation procedure so that for $j$ odd we use the circuit in Fig.~\ref{fig:3} but for $j$ even we use the alternative circuit in Fig.~\ref{fig:10}; in the latter, the direction of the {\sc cnot} gates is reversed so that the values of $c_{1{\rm x}}$ and $c_{1{\rm z}}$ are switched compared to the former, and similarly for the values of $c_{2{\rm x}}$ and $c_{2{\rm z}}$.

Finally, we observe that the recursion equations may yield a smaller estimate of $\varepsilon^{\;\;}_{\rm css}(j)$ if we omit the subblock teleportations for small values of $j$. Naively, omitting the subblock teleportations is disadvantageous; it makes the level-$(j{-}1)$ subblocks of the $j$-BP  noisier than the input $(j{-}1)$-BPs. But including the subblock teleportations means that the $j$-BP is less likely to be accepted, so that our estimate of the noise strength in the postselected $j$-BP may be higher. By analyzing several different strategies, we find that it is beneficial to omit the subblock teleportations for $j=2$, and include them for $j\geq 3$.      

\begin{figure}[t]
\begin{center}
\includegraphics[width=8cm,keepaspectratio]{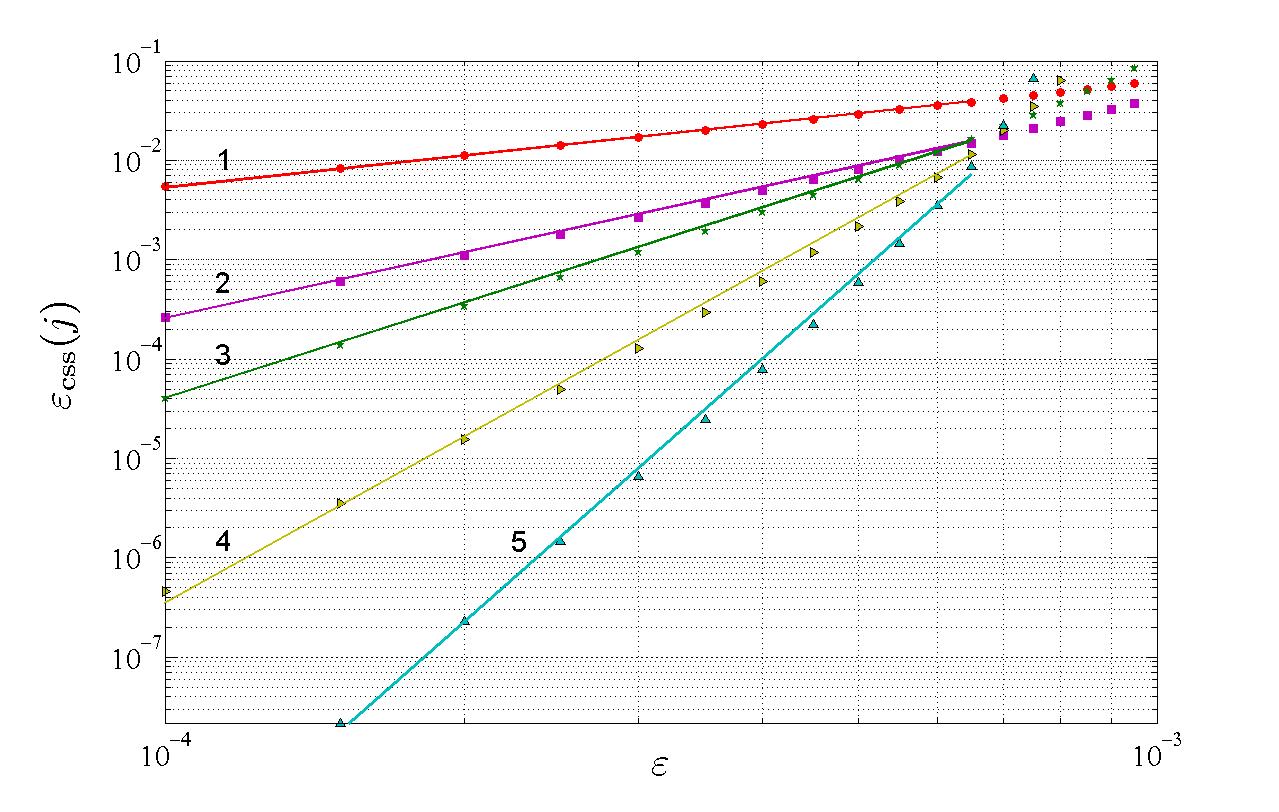}
\includegraphics[width=8cm,keepaspectratio]{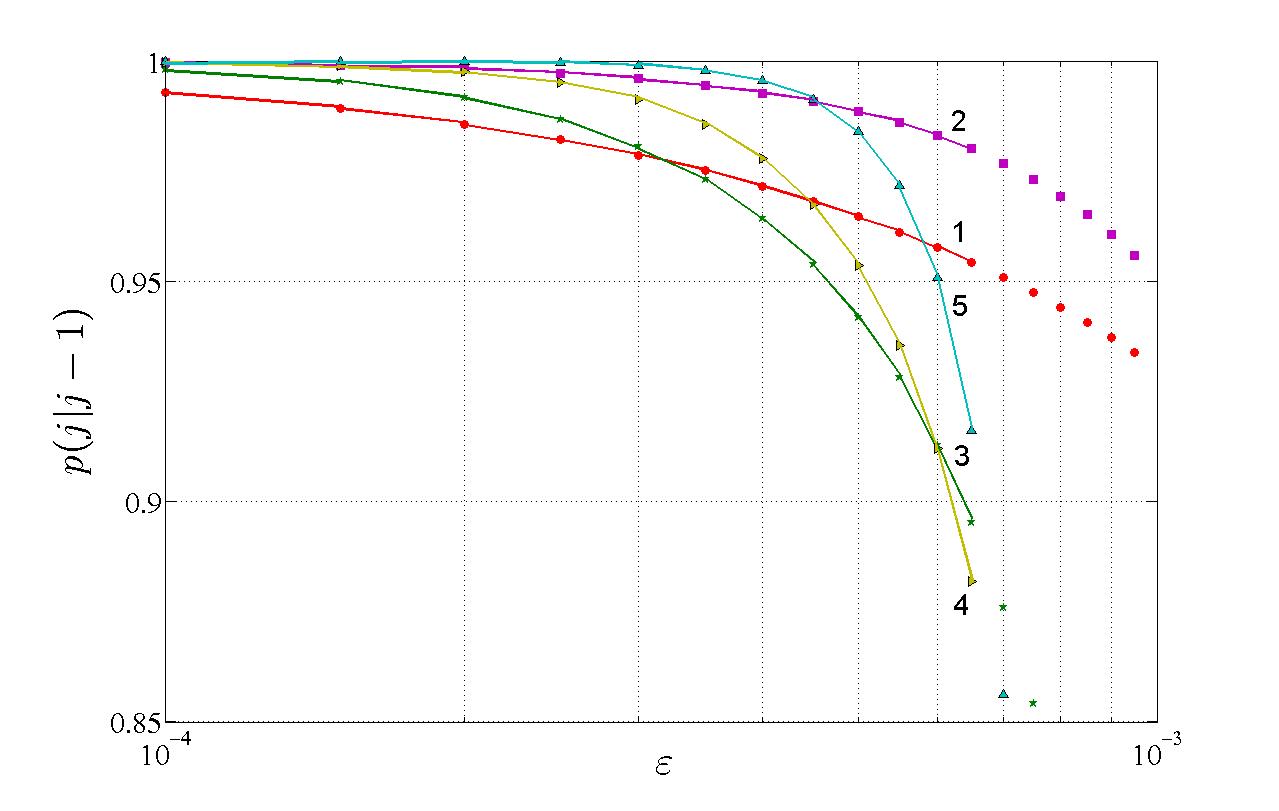}
\vspace{-1cm} 
\end{center}
\caption{\label{fig:9} On the top, the effective noise strength $\varepsilon_{\rm css}(j)$ for $\mathcal{G}_{\rm CSS}$ operations protected by $(C_4)^{\triangleright \, j}$, as a function of the fundamental noise strength $\varepsilon$, for $j=1,2,3,4,5$. The straight lines serve as guides to the eye for points with the same value of $j$. For $\varepsilon$ satisfying Eq.~(\ref{eq:threshold}), $\varepsilon_{\rm css}(j)$ decreases as $j$ increases.  On the bottom, the probability $p(j|j{-}1)$ of accepting a $j$-BP conditioned on having accepted all $(j{-}1)$-BPs, as a function of the fundamental noise strength $\varepsilon$, for $j=1,2,3,4,5$. For $\varepsilon$ satisfying Eq.~(\ref{eq:threshold}), $p(j|j{-}1)$ approaches unity as $j$ increases. The different behavior of the probability of acceptance for $j=1,2$ compared to $j\ge 3$ is due to omitting the subblock teleportations for $j=2$ but including them for $j\ge 3$.
         }
\end{figure}

Analyzing our recursion equations with these modifications, we find the results that are plotted in Fig.~\ref{fig:9}; in particular, we find that when
\begin{equation}
\label{eq:threshold}
\varepsilon\leq \varepsilon^{\rm css}_0 \equiv .67\times 10^{-3} \; ,
\end{equation} 
\noindent $\varepsilon^{\;\;}_{\rm css}(j)$ decreases with increasing $j$, thus indicating that $\varepsilon^{\rm css}_0$ is a lower bound on the accuracy threshold for $\mathcal{G}_{\rm CSS}$ circuits. It is possible by using our recursion equations to prove rigorously that $\varepsilon^{\;\;}_{\rm css}(j)$ really does become arbitrarily small for sufficiently large $j$. But a simpler method for proving the $\mathcal{G}_{\rm CSS}$ threshold estimate is to note that for $\varepsilon\leq \varepsilon^{\rm css}_0$ and $j=10$ we have $\varepsilon^{\;\;}_{\rm css}(10) \leq 1.43\times 10^{-5}$, which is smaller than the lower bound $1.26\times 10^{-4}$ on the $\mathcal{G}_{\rm CSS}$ threshold proved previously in  \cite{Aliferis06c}, using the concatenated 9-qubit Bacon-Shor code. 

In addition, Fig.~\ref{fig:9} shows the probability $p(j|j{-}1)$ of accepting a $j$-BP conditioned on the acceptance of all input $(j{-}1)$-BPs. When the threshold condition in Eq.~(\ref{eq:threshold}) is satisfied, $p(j|j{-}1)$ converges to unity with increasing $j$,  showing that the overhead cost of using postselection is moderate; the cost is dominated by the overhead for preparing $j$-BPs for small values of $j$, while for large $j$ the $j$-BPs are rarely rejected.

%---------------------------------------------------%

\subsection{The $\mathcal{G}_{\rm CSS}$ threshold for independent depolarizing noise}

We can push the $\mathcal{G}_{\rm CSS}$ accuracy threshold higher if we make further assumptions about the noise model. In particular, we note that our circuits are constructed so that errors acting on the two qubits in a {\sc cnot} gate always propagate to measurements in {\em different} blocks. We may define $\varepsilon_1\le \varepsilon$ by requiring that for either qubit in each {\sc cnot} gate, the probability of a type-m error is at most $\varepsilon_1$, for both m = x and m = z. If in addition the probability of a type-m error is at most $\varepsilon_1$ for both preparations and measurements, then we can replace $\varepsilon$ by $\varepsilon_1$ in the recursion equations stated in Tables~\ref{table:1}, \ref{table:2}, and \ref{table:3}. Thus the $\mathcal{G}_{\rm CSS}$ threshold condition becomes $\varepsilon_1 \le .67 \times 10^{-3}$.

More precisely, the noisy fundamental operation $\mathcal{O}'$ that simulates the ideal operation $\mathcal{O}$ can be expressed as $\mathcal{O}' = \mathcal{N} \circ \mathcal{O}$, where $\mathcal{N}$ is a completely-positive trace-preserving map with Kraus operators $\{N_k\}$.  In the case of the {\sc cnot} gate, for each qubit $q\in\{1,2\}$ and error type ${\rm m}\in\{{\rm x,z}\}$, we decompose the Kraus operators into two sets $\{ N_k \} = \{ N_k^{(q,{\rm m})} \} \cup \{ N_k^{({\rm rest})} \}$, where Kraus operators in the first subset do not commute with $\sigma_{\neg{\rm m}}^{(q)}$ acting on qubit $q$, and Kraus operators in the second set do commute with $\sigma_{\neg{\rm m}}^{(q)}$ (here, $\neg{\rm m}$ denotes the complement of m). In our refined noise model, the total noise strength of the Kraus operators in the set $\{ N_k^{(q,{\rm m})} \}$ is at most $\varepsilon_1$ for each $q$ and m. The single-qubit preparations and measurements obey a similar requirement.
 
A special case, {\em independent depolarizing noise}, was used in Knill's numerical simulations \cite{Knill05}. For the {\sc cnot} gate, the nontrivial Kraus operators are the 15 nontrivial Pauli operators, each occurring with probability $\varepsilon/15$. The 8 Kraus operators $\{\sigma_{\rm x},\sigma_{\rm y}\}\otimes \{I, \sigma_{\rm x},\sigma_{\rm y},\sigma_{\rm z}\}$ induce type-x errors on the first qubit; similarly, 8 Kraus operators induce type-m errors on qubit $q$ for each m and $q$. Thus, the probability of a type-m error acting on qubit $q$ is ${8\over 15}\varepsilon$ for each m and $q$. If we also assume that for each m the probability of a type-m error is at most ${8\over 15}\varepsilon$ in qubit preparations and measurements, then $\varepsilon_1 = {8\over 15}\varepsilon$. We conclude that the $\mathcal{G}_{\rm CSS}$ threshold condition is
\begin{equation}
\label{eq:threshold-depol}
\varepsilon \leq \varepsilon^{\rm css,depol}_0 = \frac{15}{8}\cdot \varepsilon^{\rm css}_0 = 1.25\times 10^{-3} \; 
\end{equation}   
\noindent for the case of independent depolarizing noise.

\begin{figure}[t]
\begin{center}
\begin{tabular}{c}
%\put(-3.8,0){
\put(-1.3,0){
\includegraphics[width=9.5cm,keepaspectratio]{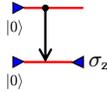} 
         }
\end{tabular}
\vspace{-6.3cm}
\end{center}
\caption{\label{fig:11} A purification circuit for the preparation of the state $|0\rangle$: Two copies of $|0\rangle$ are prepared, a {\sc cnot} gate is applied, and $\sigma_{\rm z}$ is measured on the second output qubit; the preparation is accepted if the measurement outcome is $\sigma_{\rm z}{=}+1$. For independent depolarizing noise, the probability of a type-x error conditioned on accepting the $|0\rangle$ state is ${8\over 15}\varepsilon+O(\varepsilon^2)$. %(b) A circuit for improving the reliability of the measurement of $\sigma_{\rm z}$ (the time-reversed version of the circuit on the left): An ancilla is prepared in the state $|0\rangle$, a {\sc cnot} gate is applied, and $\sigma_{\rm z}$ is measured on both qubits; the measurement is accepted if the measurement outcomes agree, otherwise the state is rejected. For independent depolarizing noise, the probability of error conditioned on acceptance is ${4\over 15}\varepsilon +O(\varepsilon^2)$. 
         }
\end{figure}

For the independent depolarizing noise model, it might seem natural to assume that the probability of a type-m error in single-qubit operations such as qubit preparations and measurements is $\frac{2}{3}\varepsilon$, since two of the three nontrivial single-qubit Pauli operators induce such an error. To simplify our calculations, we have assumed that the error probability is ${8\over 15}\varepsilon$ instead. One way to defend this assumption is to note that the reliability of the preparation of the state $|0\rangle$ can be improved using the purification circuit shown in Fig.~\ref{fig:11}; a similar circuit can be used to purify the preparation of the state $|+\rangle$. The preparation of $|0\rangle$ in Fig.~\ref{fig:11} is accepted only if the measurement outcome is $\sigma_{\rm z}{=}+1$. Conditioned on accepting the $|0\rangle$ state, then, a type-x error with $O(\varepsilon)$ probability can arise only from a fault in the {\sc cnot} gate, not from a single fault in one of the fundamental preparation steps. Thus the conditional probability of a type-x error in the accepted state is ${8\over 15}\varepsilon+O(\varepsilon^2)$. 

%The reliability of the measurement of $\sigma_{\rm z}$ can be improved using the circuit shown in Fig.~\ref{fig:11}(b); a similar circuit can be used for the measurement of $\sigma_{\rm x}$. The $\sigma_{\rm z}$ measurement in Fig.~\ref{fig:11}(b) is accepted only if both measurement outcomes agree; otherwise the state is rejected. Conditioned on acceptance, a type-x error with $O(\varepsilon)$ probability can arise only from a fault in the {\sc cnot} gate, not from a single fault in one of the fundamental measurement steps. Only the 4 Pauli operators $\{\sigma_{\rm x},\sigma_{\rm y}\}\otimes \{\sigma_{\rm x},\sigma_{\rm y}\}$ in the noisy {\sc cnot} gate induce accepted type-x errors. Thus the conditional probability of a type-x error in an accepted measurement is ${4\over 15}\varepsilon+O(\varepsilon^2)$. %But this postselection affects our estimates on the failure probabilities for the measured blocks.

%---------------------------------------------------%

\section{Universality}
\label{sec:threshold}

Now we have shown that, if $\varepsilon$ is sufficiently small, $\mathcal{G}_{\rm CSS}$ gates can be simulated with arbitrarily good accuracy. To complete our proof of a quantum accuracy threshold theorem based on the Fibonacci scheme, we must extend this protected gate set to a universal set. It suffices to augment the highly reliable encoded $\mathcal{G}_{\rm CSS}$ gates with highly reliable encoded versions of the single-qubit preparations $\mathcal{P}_{|{+}i\rangle}$ and $\mathcal{P}_{|T\rangle}$. 

Using the reliable encoded $\mathcal{G}_{\rm CSS}$ operations, high-fidelity encoded copies of $|{+}i\rangle$ and $|T\rangle$ can be {\em distilled} starting with sufficiently accurate noisy copies. The distillation protocol works if encoded copies of $|T\rangle$ can be prepared such that the probability of a logical error is below $14.1\%$ \cite{Bravyi04}; the distillation threshold for $|{+}i\rangle$ states is even higher \cite{Aliferis07b}. 

One way to prepare encoded copies of $|\psi\rangle$  is by ``teleportation into the code block'' \cite{Knill05}. First, we prepare a Bell pair encoded using $(C_4)^{\triangleright \, j}$, then we recursively decode one of the two encoded blocks to a single qubit, and finally we perform a Bell measurement on the decoded qubit and an input qubit prepared in the state $|\psi\rangle$. The level-$j$ encoded Bell pair can be prepared reliably if $\varepsilon$ is below the $\mathcal{G}_{\rm CSS}$ threshold and $j$ is large. But the teleportation procedure could fail either because there is an error in the recursively decoded state, or because a fault occurs in the preparation of $|\psi\rangle$ or in the two-qubit Bell-measurement circuit.

Let us consider the probability of error in the recursive decoding of a $(C_4)^{\triangleright \, j}$ block to a single qubit. A level-1 $C_4$ block can be decoded to a qubit using a circuit that contains three {\sc cnot} gates \cite{Aliferis07b}. In the first decoding step we use level-$(j{-}1)$ {\sc cnot} gates to decode the $(C_4)^{\triangleright \, j}$ block to a $(C_4)^{\triangleright \, (j{-}1)}$ block, in the second decoding step we use level-$(j{-}2)$ {\sc cnot} gates to decode the $(C_4)^{\triangleright \, (j{-}1)}$ block to a $(C_4)^{\triangleright \, (j{-}2)}$ block, and so on, completing the decoding to a single qubit after $j$ steps.

Each encoded {\sc cnot} gate is followed by error-correcting teleportation steps acting on both output blocks. We abort the ``teleportation into the block'' protocol if, at any stage of decoding the $(C_4)^{\triangleright \, j}$ block, a flag is raised in the top-level decoding of a measurement inside any error-correcting teleportation.  (We assume that sufficiently many encoded Bell pairs are prepared in parallel that with high probability at least one ancilla preparation is accepted.) A flag could be raised in either the $\sigma_{\rm x}$ or $\sigma_{\rm z}$ measurement for either output block of any {\sc cnot} gate at any level, so the total probability of a flag during decoding is
\begin{equation}
f_{\rm dec}(j) \leq 3\sum\limits_{r\in\{{\rm c},{\rm t} \}} \sum\limits_{{\rm m}\in \{{\rm x}, {\rm z} \}} \sum\limits_{i=1}^{j-1} f^r_{\sigma_{\rm m}}(i,i|i) \; .
\end{equation}
\noindent A logical error, too, could occur in either the $\sigma_{\rm x}$ or $\sigma_{\rm z}$ measurement for either output block of any {\sc cnot} gate at any level, or it might occur in one of the three level-0 {\sc cnot} gates in the final stage of decoding, so the joint probability of acceptance and a logical error in the decoded qubit is 
\begin{widetext}
\begin{equation}
\label{eq:decoding-error}
\varepsilon_{\rm dec}(j) \leq 3 \sum\limits_{r\in\{{\rm c},{\rm t} \}} \sum\limits_{{\rm m}\in \{{\rm x}, {\rm z} \}} \sum_{i=1}^{j-1} r_{\sigma_{\rm m}}(i,i{\wedge}\neg f|i) + 3\, \varepsilon\; .
\end{equation} 
\end{widetext}
\noindent We find then, that the probability of failure for the teleportation into the code block is  
\begin{equation}
\varepsilon_{\rm anc}^L(j) \leq 4\varepsilon + {\varepsilon_{\rm dec}(j) \over 1-f_{\rm dec}(j)} \; ;
\end{equation}
the second term is an upper bound on the probability of error in the decoded qubit conditioned on acceptance, and the first term is an upper bound on the probability of a fault in the level-0 teleportation circuit (the $|\psi\rangle$ preparation, one {\sc cnot} gate, and two single-qubit measurements).

It is possible by using our recursion equations to prove that if the fundamental noise strength $\varepsilon$ is below the  $\mathcal{G}_{\rm CSS}$ accuracy threshold, then, for all $j$, $\varepsilon_{\rm anc}^L(j)$ is bounded above by $6.1\%$ for local stochastic noise and $6.8\%$ for independent depolarizing noise. Thus, in both cases encoded ancilla states can be prepared with a probability of error below the distillation threshold, and robust universal quantum computation is possible. Alternatively, we can bypass the need to bound $\varepsilon_{\rm anc}^L(j)$ for all $j$  by noting that for $j=10$ we have $\varepsilon_{\rm anc}^L(10)\leq 6.09\%$ for local stochastic noise and $\varepsilon_{\rm anc}^{L,{\rm depol}}(10)\leq 6.76\%$ for independent depolarizing noise. It was proven previously that, for a scheme based on concatenation of the 9-qubit Bacon-Shor code $C_9$, if the strength of local stochastic noise is below the $\mathcal{G}_{\rm CSS}$ threshold, then for any $k$ a $(C_9)^{\triangleright \, k}$ block can be decoded with error probability at most $2.15\%$ \cite{Aliferis07}. Therefore we can concatenate the Fibonacci scheme at level $j=10$ with the Bacon-Shor scheme, and perform teleportation into the code block with a total probability of a logical error no larger than $8.24\%$ for local stochastic noise and $8.91\%$ for independent depolarizing noise, still below the distillation threshold. 

We conclude, then, that for the Fibonacci scheme the $\mathcal{G}_{\rm CSS}$ accuracy threshold is also a threshold for universal quantum computation. In summary, we have proved:

\begin{thm} [Quantum accuracy threshold theorem for the Fibonacci scheme] \label{theorem:1} An ideal quantum circuit can be accurately simulated by a noisy quantum circuit using the Fibonacci scheme for fault-tolerant quantum computation if the fundamental noise strength $\varepsilon$ is no larger than $\varepsilon_0$, where $\varepsilon_0= .67\times 10^{-3}$ for local stochastic noise and $\varepsilon_0= 1.25\times 10^{-3}$ for independent depolarizing noise. The ratio of the size $L^*$ of the noisy circuit to the size $L$ of the ideal circuit is bounded above by a polynomial in the logarithm of $L$ as specified in section \ref{sec:overhead} below.
\end{thm}

%---------------------------------------------------%
\section{Overhead}
\label{sec:overhead}

The overhead factor $L^*/L$ is polylogarithmic in $L$ because, although the number of locations inside a level-$j$ gadget grows approximately exponentially with the level of concatenation $j$, the probability of a logical error in a level-$j$ gadget decreases {\em double exponentially} with $j$, if $\varepsilon$ is less than the threshold value $\varepsilon_0$.

For the Fibonacci scheme, the probability of a (flagged) logical error at level $j$ scales with $j$ according to Eq.~(\ref{flagged-failure-scaling}):
\begin{equation}
\delta_f^{(j)} \sim  \varepsilon_0 \left(\varepsilon/\varepsilon_0\right)^{F(j)}~.
\end{equation}
\noindent We can simulate a circuit of size $L$ with reasonable accuracy if the probability of a logical error per gate is of order $1/L$, or 
\begin{equation}
F(j) \approx \frac{\log(L\varepsilon_0)}{\log(\varepsilon_0/\varepsilon)}~.
\end{equation}
\noindent If the size of the largest level-$j$ gadget is bounded above by $\ell^j$ (where $\ell$ is a constant), then the overhead factor can be expressed as
\begin{equation}
\begin{array}{rcl}
L^*/L & \le & \ell^j = \left(F(j)\right)^{\log \ell^j/\log F(j)} \\
      & \approx & \left(\frac{\log(L\varepsilon_0)}{\log(\varepsilon_0/\varepsilon)}\right)^{\log \ell/\log \phi}~.
\end{array}
\end{equation}
\noindent This is polylogarithmic scaling as asserted in Theorem 1.

To examine in more detail how the gadget size depends on $j$, we must take into account that our procedure for preparing a $j$-BP uses {\em postselection}: The $j$-BP is rejected if a flag is raised at the top level in either one of the two block teleportations or in any one of the eight subblock teleportations. Thus the probability of rejection scales with $j$ like the probability of a top-level flag in a level-$(j{-}1)$ block, which is $O(\varepsilon^{F(j{-}1)})$ according to Eqs.~(\ref{flag-strength-scaling}) and (\ref{flagged-failure-scaling}). For the purpose of estimating the overhead, we may imagine that a constant number (independent of $j$) of $j$-BP preparations are executed in parallel, where the constant is chosen to be large enough so that the probability that all of the $j$-BPs are rejected is sufficiently small. 

For example, for each $j$-BP used in the level-$j$ {\sc cnot} gadget, suppose that $M$ $j$-BPs are prepared in parallel. The {\sc cnot} gadget may fail because, in one of the $j$-BP preparations contained inside the gadget, all $M$ of the $j$-BPs are rejected---this occurs with probability $O(\varepsilon^{MF(j{-}1)})$. The probability of a logical error in the {\sc cnot} gadget arising from other sources is $O(\varepsilon^{F(j)})$; thus the probability of failure due to the rejection of all $M$ $j$-BPs is comparable to other contributions to the failure probability if we choose
\begin{equation}
M \approx {F(j)\over F(j{-}1)} \approx \phi\approx 1.618 \; .
\end{equation}

Similarly, for each input $j$-BP used in the preparation of a level-$(j{+}1)$ BP, suppose that $N$ $j$-BPs are prepared in parallel. Then the probability that all $N$ $j$-BPs are rejected is $O(\varepsilon^{NF(j{-}1)})$. In this case, $N$ should be large enough that the probability of failure due to rejection of all the $j$-BPs is comparable to the probability of an unflagged level-$j$ logical error in an accepted $j$-BP, which is $O(\varepsilon^{F(j{+}1)})$ according to Eq.~(\ref{unflagged-error-scaling}). Therefore we choose
\begin{equation}
N \approx {F(j+1)\over F(j{-}1)} \approx \phi^2\approx 2.618 \; .
\end{equation}

The $j$-BP preparation circuit (for $j>1$) shown in Fig.~\ref{fig:3} uses as input 12 $(j{-}1)$-BPs, plus another 8 $(j{-}1)$-BPs for the final subblock teleportations. For our overhead estimate, we replace these 20 $(j{-}1)$-BP preparations by $20N$ $(j{-}1)$-BP preparations. In addition, the $j$-BP preparation circuit contains 28 level-$(j{-}1)$ transversal logical {\sc cnot} gates ($12$ to prepare three level-$j$ Bell pairs, $8$ for the two block teleportations, and $8$ for the eight subblock teleportations) plus 32 level-$(j{-}1)$ transversal measurements ($16$ for the two block teleportations and $16$ for the eight subblock teleportations). 

More generally, suppose that the preparation circuit for a $j$-BP ($j > 1$) uses $r$ $(j{-}1)$-BPs,  $s$ level-$(j{-}1)$ {\sc cnot} gates and $t$ level-$(j{-}1)$ measurements. Furthermore, suppose that the logical {\sc cnot} gates and measurements are transversal, so that a level-$j$ logical {\sc cnot} uses $n^j$ fundamental two-qubit {\sc cnot} gates and a level-$j$ logical measurement uses $n^j$ fundamental single-qubit measurements, where $n$ is the block size of the code that is concatenated $j$ times ($n\,{=}\,4$ for the scheme we have analyzed). Let $B(j)$ denote the number of $1$-BP preparations, $C(j)$ denote the number of level-1 {\sc cnot} gates, and $M(j)$ denote the number of level-1 measurements contained in a level-$j$ BP preparation. These quantities obey the recursion relations:
\begin{equation}
\begin{array}{rcl}
B(j) & = & r B(j{-}1) ~, \\
C(j) & = & r C(j{-}1) + s n^{j{-}2}~, \\
M(j) & = & r M(j{-}1) + t n^{j{-}2} ~.
\end{array}
\end{equation}
\noindent Solving these recursion relations, we find 
\begin{equation}
\begin{array}{rcl}
B(j) & = & r^{j{-}1} ~, \\
C(j) & = & s\left( r^{j{-}1} - n^{j{-}1}\right)/(r-n)~, \\
M(j) & = & t\left( r^{j{-}1} - n^{j{-}1}\right)/(r-n) ~,
\end{array}
\end{equation}
\noindent assuming $r\ne n$. These expressions illustrate how the number of locations inside a level-$j$ gadget grows roughly exponentially with $j$, at least for the case of the $\mathcal{G}_{\rm CSS}$ gadgets. The distillation protocols that prepare high-fidelity encoded ancillas for non-Clifford gates use $\mathcal{G}_{\rm CSS}$ gadgets, and require a number of rounds of that grows linearly with $j$; thus the circuit size for the non-Clifford gadgets can also be bounded above by an exponential in $j$. 

An important feature of the (non-recursive) Fibonacci scheme, not shared by the strictly recursive schemes analyzed in \cite{Aliferis05b}, is that although the size $L^*$ of the noisy circuit is larger than the size $L$ of the ideal circuit by a factor polylogarithmic in $L$, the ``quantum depth'' $D^*$ of the noisy circuit is larger than the depth $D$ of the ideal circuit by only a constant factor. (That is, the number of time steps in the simulation is larger than the number of time steps in the ideal computation by a constant factor, {\em if} we neglect the time required for classical processing of measurement outcomes.) 

Note that the preparation of Bell pairs and the distillation of ancilla states are performed off-line and hence do not contribute to the simulation's depth. A transversal level-$j$ logical {\sc cnot} gate can be executed in a single time step for any value of $j$. The quantum measurements comprising the transversal measurement of a level-$j$ block can also be performed in a single time step, but to determine the outcome of the encoded measurement classical processing of the measurement outcomes is required, which can be executed by a classical circuit with depth linear in $j$. 

Whenever a non-Clifford gate is teleported, the outcome of the encoded measurement is needed to determine what Clifford gate should be applied in the next step. If classical processing is much faster than quantum processing, then we may neglect the time needed for the classical decoding of measured blocks; under this assumption, non-Clifford gates as well as Clifford gates can be simulated in constant depth.

We will not present here a more detailed analysis of the resource requirements for the Fibonacci scheme, but we note that estimates based on numerical simulations are discussed in the Supplementary Information in Ref.~\cite{Knill05}.

%---------------------------------------------------%
\section{Conclusion}
\label{sec:conclusion}

We can expect the accuracy of quantum computing hardware to improve steadily. Even so, overcoming the limiting effects of noise will pose a challenge to the builders of quantum computers well into the future, and  fault-tolerant simulations will be needed to operate large-scale quantum computers reliably. There are two central questions about fault-tolerant schemes: how much noise can be tolerated and how does the overhead cost depend on the strength of the noise?

Much can be learned about the answers to these questions through numerical studies of fault-tolerant circuits. But it is also useful to prove rigorous statements about the performance of fault-tolerant methods. For one thing, a long computation might fail due to rare noise fluctuations; the probability of rare events could be seriously underestimated in numerical simulations, so that rigorous upper bounds on the probability are valuable. Furthermore, rigorous analysis can deepen our conceptual understanding regarding which approaches to fault tolerance are most effective, and inspire new proposals for fault-tolerant circuit design.

This paper, like the earlier paper \cite{Aliferis07b}, provides a rigorous analysis of a clever approach to quantum fault tolerance introduced by Knill \cite{Knill05}; in \cite{Aliferis07b} we analyzed Knill's ``postselection scheme'' and here we have analyzed his ``Fibonacci scheme,'' which has a much more favorable overhead cost than the postselection scheme. In each case we have established a rigorous lower bound on the quantum accuracy threshold that is significantly higher than the previously proven lower bounds for other fault-tolerant schemes, though still about an order of magnitude below Knill's numerical estimates. (High numerical estimates of the accuracy threshold have also been found in \cite{raussendorf}, though for that scheme a fully rigorous analysis has not yet been done.)

For the Fibonacci scheme, our rigorous lower bound on the threshold noise strength $\varepsilon_0$ is $.67\times 10^{-3}$ for local stochastic noise and $1.25\times 10^{-3}$ for independent depolarizing noise. Our previously derived lower bound for the postselection scheme was $1.04 \times 10^{-3}$, but that result applies to a modified noise model with weaker noise correlations than local stochastic noise and stronger noise correlations than independent depolarizing noise. All of these results notably improve on the lower bound $1.94\times 10^{-4}$ for local stochastic noise found earlier for the concatenated Bacon-Shor scheme \cite{Aliferis06c}.

Our proof of the threshold theorem uses some new methods that might be fruitfully applied elsewhere. We have introduced a new wrinkle on ``level reduction'' by characterizing the noise in recursively prepared Bell pairs using a hierarchical effective circuit that produces the same output as the actual Bell-pair preparation circuit. We have also developed tools for analyzing the recursive decoding of concatenated code blocks assisted by flags that point to probable error locations. In a separate paper, we have shown that flagging can be helpfully invoked in schemes designed to protect against highly biased noise models, where type-z errors are far more likely than  type-x errors \cite{Aliferis07c}. 

This paper supersedes an earlier discussion \cite{fibonacci-old} where a recursive scheme which used flagging was studied. The analysis reported in \cite{fibonacci-old} was flawed, and the threshold estimate derived there was incorrect. 

%---------------------------------------------------%
%\section{Acknowledgments}
\acknowledgments
We are grateful to Daniel Gottesman, David DiVincenzo, and Barbara Terhal for helpful discussions and comments. This research is supported in part by DoE under Grant No. DE-FG03-92-ER40701, NSF under Grant No. PHY-0456720, and NSA under ARO Contract No. W911NF-05-1-0294.

%-----------------------------------------------------------------------------------------------------------%
\bibliographystyle{unsrt}

\end{document}